# Packaging research artefacts with RO-Crate

Stian Soiland-Reyes [a,b,*], Peter Sefton [c], Mercè Crosas [d], Leyla Jael Castro [e], Frederik Coppens [f], José M. Fernández [g], Daniel Garijo [h], Björn Grüning [i], Marco La Rosa [j], Simone Leo [k], Eoghan Ó Carragáin [l], Marc Portier [m], Ana Trisovic [n], RO-Crate Community [o], Paul Groth [p], Carole Goble [q]

[a] *Department of Computer Science, The University of Manchester, UK*
E-mail: *soiland-reyes@manchester.ac.uk*; ORCID: *https://orcid.org/0000-0001-9842-9718*
[b] *Informatics Institute, University of Amsterdam, The Netherlands*
[c] *Faculty of Science, University Technology Sydney, Australia*
E-mail: *Peter.Sefton@uts.edu.au*; ORCID: *https://orcid.org/0000-0002-3545-944X*
[d] *Institute for Quantitative Social Science, Harvard University, Cambridge, MA, USA*
ORCID: *https://orcid.org/0000-0003-1304-1939*
[e] *ZB MED Information Centre for Life Sciences, Cologne, Germany*
ORCID: *https://orcid.org/0000-0003-3986-0510*
[f] *VIB-UGent Center for Plant Systems Biology, Gent, Belgium*
ORCID: *https://orcid.org/0000-0001-6565-5145*
[g] *Barcelona Supercomputing Center, Barcelona, Spain*
ORCID: *https://orcid.org/0000-0002-4806-5140*
[h] *Ontology Engineering Group, Universidad Politécnica de Madrid, Madrid, Spain*
ORCID: *https://orcid.org/0000-0003-0454-7145*
[i] *Bioinformatics Group, Department of Computer Science, Albert-Ludwigs-University Freiburg, Freiburg, Germany*
ORCID: *https://orcid.org/0000-0002-3079-6586*
[j] *PARADISEC, Melbourne, Australia*
ORCID: *https://orcid.org/0000-0001-5383-6993*
[k] *Center for Advanced Studies, Research, and Development in Sardinia (CRS4), Pula (CA), Italy*
ORCID: *https://orcid.org/0000-0001-8271-5429*
[l] *University College Cork, Ireland*
ORCID: *https://orcid.org/0000-0001-8131-2150*
[m] *Vlaams Instituut voor de Zee, Oostende, Belgium*
ORCID: *https://orcid.org/0000-0002-9648-6484*
[n] *Institute for Quantitative Social Science, Harvard University, Cambridge, MA, USA*
ORCID: *https://orcid.org/0000-0003-1991-0533*
[o] *https://www.researchobject.org/ro-crate/community* (see appendix B)
[p] *Informatics Institute, University of Amsterdam, The Netherlands*
E-mail: *p.t.groth@uva.nl*; ORCID: *https://orcid.org/0000-0003-0183-6910*
[q] *Department of Computer Science, The University of Manchester, UK*

E-mail: carole.goble@manchester.ac.uk; ORCID: https://orcid.org/0000-0003-1219-2137

**Abstract.** An increasing number of researchers support reproducibility by including pointers to and descriptions of datasets, software and methods in their publications. However, scientific articles may be ambiguous, incomplete and difficult to process by automated systems. In this paper we introduce RO-Crate, an open, community-driven, and lightweight approach to packaging research artefacts along with their metadata in a machine readable manner. RO-Crate is based on Schema.org annotations in JSON-LD, aiming to establish best practices to formally describe metadata in an accessible and practical way for their use in a wide variety of situations.

An RO-Crate is a structured archive of all the items that contributed to a research outcome, including their identifiers, provenance, relations and annotations. As a general purpose packaging approach for data and their metadata, RO-Crate is used across multiple areas, including bioinformatics, digital humanities and regulatory sciences. By applying "just enough" Linked Data standards, RO-Crate simplifies the process of making research outputs FAIR while also enhancing research reproducibility.

An RO-Crate for this article[1] is archived at `https://doi.org/10.5281/zenodo.5146227`

## 1. Introduction

The move towards Open Science has increased the need and demand for the publication of artefacts of the research process [1]. This is particularly apparent in domains that rely on computational experiments; for example, the publication of software, datasets and records of the dependencies that such experiments rely on [2].

It is often argued that the publication of these assets, and specifically software [3], workflows [4] and data, should follow the FAIR principles [5]; namely, that they are Findable, Accessible, Interoperable and Reusable. These principles are agnostic to the *implementation* strategy needed to comply with them. Hence, there has been an increasing amount of work in the development of platforms and specifications that aim to fulfil these goals [6].

Important examples include data publication with rich metadata (e.g. Zenodo [7]), domain-specific data deposition (e.g. PDB [8]) and following practices for reproducible research software [9] (e.g. use of containers). While these platforms are useful, experience has shown that it is important to put greater emphasis on the interconnection of the multiple artefacts that make up the research process [10].

The notion of **Research Objects** [11] (RO) was introduced to address this connectivity, providing semantically rich *aggregations* of (potentially distributed) resources with a layer of structure over a research study; this is then to be delivered in a *machine-readable format*.

A Research Object combines the ability to bundle multiple types of artefacts together, such as spreadsheets, code, examples, and figures. The RO is augmented with annotations and relationships that describe the artefacts' *context* (e.g. a CSV being used by a script, a figure being a result of a workflow).

This notion of ROs provides a compelling vision as an approach for implementing FAIR data. However, existing Research Object implementations require a large technology stack [12], are typically tailored to a particular platform and are also not easily usable by end-users.

---

[*]Corresponding author. E-mail: soiland-reyes@manchester.ac.uk.
[1]<https://w3id.org/ro/doi/10.5281/zenodo.5146227>

To address this gap, a new community came together [13] to develop **RO-Crate** — an *approach to package and aggregate research artefacts with their metadata and relationships*. The aim of this paper is to introduce RO-Crate and assess it as a strategy for making multiple types of research artefacts FAIR. Specifically, the contributions of this paper are as follows:

1. An introduction to RO-Crate, its purpose and context;
2. A guide to the RO-Crate community and tooling;
3. Examples of RO-Crate usage, demonstrating its value as connective tissue for different artefacts from different communities.

The rest of this paper is organised as follows. We first describe RO-Crate through its development methodology that formed the RO-Crate concept, showing its foundations in Linked Data and emerging principles. We then define RO-Crate technically, before we introduce the community and tooling. We move to analyse RO-Crate with respect to usage in a diverse set of domains. Finally, we present related work and conclude with some remarks including RO-Crate highlights and future work. The appendix adds a formal definition of RO-Crate using First-Order logic.

**2. RO-Crate**

RO-Crate aims to provide an approach to packaging research artefacts with their metadata that can be easily adopted. To illustrate this, let us imagine a research paper reporting on the sequence analysis of proteins obtained from an experiment on mice. The sequence output files, sequence analysis code, resulting data and reports summarising statistical measures are all important and inter-related research artefacts, and consequently would ideally all be co-located in a directory and accompanied with their corresponding metadata. In reality, some of the artefacts (e.g. data or software) will be recorded as external reference to repositories that are not necessarily following the FAIR principles. This conceptual directory, along with the relationships between its constituent digital artefacts, is what the RO-Crate model aims to represent, linking together all the elements of an experiment that are required for the experiment's reproducibility and reusability.

The question then arises as to how the directory with all this material should be packaged in a manner that is accessible and usable by others. This means programmatically and automatically accessible by machines and human readable. A de facto approach to sharing collections of resources is through compressed archives (e.g. a zip file). This solves the problem of "packaging", but it does not guarantee downstream access to all artefacts in a programmatic fashion, nor describe the role of each file in that particular research. Both features, the ability to automatically access and reason about an object, are crucial and lead to the need for explicit metadata about the contents of the folder, describing each and linking them together.

Examples of metadata descriptions across a wide range of domains[2] abound within the literature, both in research data management [14] [15] [16] and within library and information systems[3] [17] [18]. However, many of these approaches require knowledge of metadata schemas, particular annotation systems, or the use of complex software stacks. Indeed, particularly within research, these requirements have led to a lack of adoption and growing frustration with current tooling and specifications [19] [20] [21].

RO-Crate seeks to address this complexity by:

---
[2]<https://rdamsc.bath.ac.uk/scheme-index>
[3]<https://www.loc.gov/librarians/standards>

1. being conceptually simple and easy to understand for developers;
2. providing strong, easy tooling for integration into community projects;
3. providing a strong and opinionated guide regarding current best practices;
4. adopting de-facto standards that are widely used on the Web.

In the following sections we demonstrate how the RO-Crate specification and ecosystem achieve these goals.

*2.1. Development Methodology*

It is a good question as to what base level we assume for 'conceptually simple'. We take simplicity to apply at two levels: for the *developers* who produce the platforms and for the *data practitioners* and users of those platforms.

For our development methodology we followed the mantra of working closely with a small group to really get a deep understanding of requirements and ensure rapid feedback loops. We created a pool of early adopter projects from a range of disciplines and groups, primarily addressing developers of platforms. Thus the base level for simplicity was **developer friendliness**.

We assumed a developer familiar with making Web applications with JSON data (who would then learn how to make *RO-Crate JSON-LD*), which informed core design choices for our JSON-level documentation approach and RO-Crate serialization (section 2.3). Our group of early adopters, growing as the community evolved, drove the RO-Crate requirements and provided feedback through our multiple communication channels including bi-monthly meetings, which we describe in section 2.4 along with the established norms.

Addressing the simplicity of understanding and engaging with RO-Crate by data practitioners is through the platforms, for example with interactive tools (section 3) like Describo[4] [22] and Jupyter notebooks [23], and by close discussions with domain scientists on how to appropriately capture what they determine to be relevant metadata. This ultimately requires a new type of awareness and learning material separate from developer specifications, focusing on the simplicity of extensibility to serve the user needs, along with user-driven development of new RO-Crate Profiles specific for their needs (section 4).

*2.2. Conceptual Definition*

A key premise of RO-Crate is the existence of a wide variety of resources on the Web that can help describe research. As such, RO-Crate relies on the Linked Data principles [24]. Figure 1 shows the main conceptual elements involved in an RO-Crate: The RO-Crate Metadata File (top) describes the Research Object using structured metadata including external references, coupled with the contained artefacts (bottom) bundled and described by the RO-Crate.

The conceptual notion of a *Research Object* [11] is thus realized with the RO-Crate model and serialized using Linked Data constructs within the RO-Crate metadata file.

*2.2.1. Linked Data as a foundation*

The **Linked Data** principles [29] (use of IRIs[5] to identify resources (i.e. artefacts), resolvable via HTTP, enriched with metadata and linked to each other) are core to RO-Crate; therefore IRIs are used to identify an RO-Crate, its constituent parts and metadata descriptions, and the properties and classes used in the metadata.

---
[4]<https://uts-eresearch.github.io/describo/>
[5]**IRI**s [30] are a generalisation of *URI*s (which include well-known http/https URLs), permitting international Unicode characters without percent encoding, commonly used on the browser address bar and in HTML5.

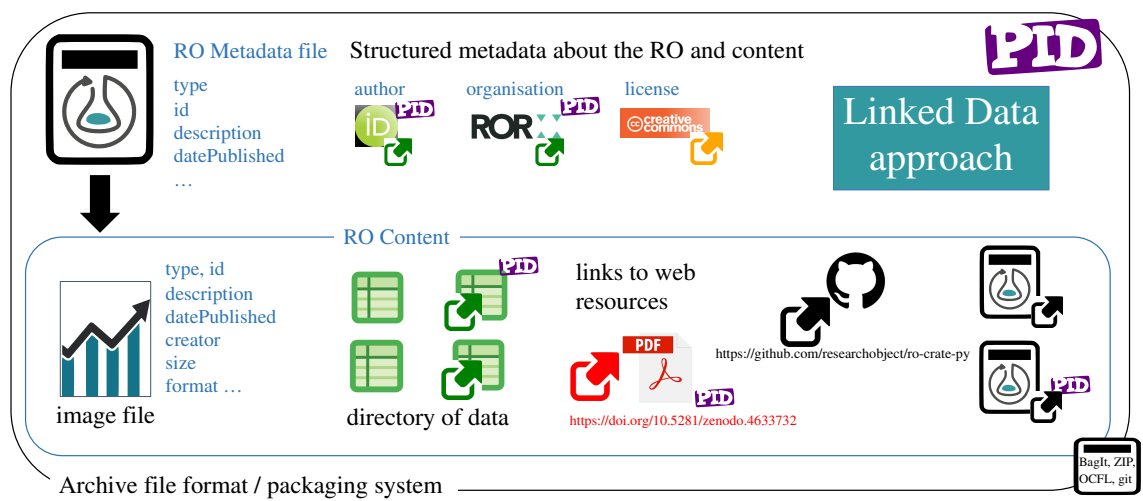

Fig. 1. **Conceptual overview of RO-Crate**. A *Persistent Identifier* (PID) [25] points to a *Research Object* (RO), which may be archived using different packaging approaches like BagIt [26], OCFL [27], git or ZIP. The RO is described within a *RO-Crate Metadata File*, providing identifiers for *authors* using ORCID, *organisations* using Research Organization Registry (ROR) [28] and licences such as Creative Commons using SPDX identifiers. The *RO-Crate content* is further described with additional metadata following a Linked Data approach. Data can be embedded files and directories, as well as links to external Web resources, PIDs and nested RO-Crates.

RO-Crates are *self-described* and follow the Linked Data principles to describe all of their resources in both human and machine readable manner. Hence, resources are identified using global identifiers (absolute IRIs) where possible; and relationships between two resources are defined with links.

The foundation of Linked Data and shared vocabularies also means that multiple RO-Crates and other Linked Data resources can be indexed, combined, queried, validated or transformed using existing Semantic Web technologies such as SPARQL[6], SHACL[7] and well established *knowledge graph* triple stores like Apache Jena[8] and OntoText GraphDB[9].

The possibilities of consuming[10] RO-Crate metadata with such powerful tools gives another strong reason for using Linked Data as a foundation. This use of mature Web[11] technologies also means its developers and consumers are not restricted to the Research Object aspects that have already been specified by the RO-Crate community, but can extend and integrate RO-Crate in multiple standardized ways.

---

[6]<https://www.w3.org/TR/sparql11-overview>

[7]<https://www.w3.org/TR/shacl/>

[8]<https://jena.apache.org/>

[9]<https://www.ontotext.com/products/graphdb/>

[10]Some consideration is needed in processing of RO-Crates as knowledge graphs, e.g. establishing absolute URIs for files inside a ZIP archive, detailed in the RO-Crate specification: https://www.researchobject.org/ro-crate/1.1/appendix/relative-uris.html

[11]Note that an RO-Crate is not required to be published on the Web, see section 2.2.2.

*2.2.2. RO-Crate is a self-described container*

An RO-Crate is defined[12] as a self-described **Root Data Entity** that describes and contains *data entities*, which are further described by referencing *contextual entities*. A **data entity** is either a *file* (i.e. a byte sequence stored on disk somewhere) or a *directory* (i.e. set of named files and other directories). A file does not need to be stored inside the RO-Crate root, it can be referenced via a PID/IRI. A **contextual entity** exists outside the information system (e.g. a Person, a workflow language) and is stored solely by its metadata. The representation of a **data entity** as a byte sequence makes it possible to store a variety of research artefacts including not only data but also, for instance, software and text.

The Root Data Entity is a directory, the *RO-Crate Root*, identified by the presence of the **RO-Crate Metadata File** `ro-crate-metadata.json` (top of Figure 1). This is a JSON-LD file that describes the RO-Crate, its content and related metadata using Linked Data. JSON-LD is a W3C standard RDF serialisation that has become popular as it is easy to read by humans while also offers some advantages for data exchange on the Internet. JSON-LD, a subset of the widely supported and well-known JSON format, has tooling available for many programming languages[13].

The minimal requirements for the root data entity metadata[14] are `name`, `description` and `datePublished`, as well as a contextual entity identifying its `license` — additional metadata are commonly added to entities depending on the purpose of the particular RO-Crate.

RO-Crates can be stored, transferred or published in multiple ways, e.g. BagIt [26], Oxford Common File Layout [27] (OCFL), downloadable ZIP archives in Zenodo or through dedicated online repositories, as well as published directly on the Web, e.g. using GitHub Pages[15]. Combined with Linked Data identifiers, this caters for a diverse set of storage and access requirements across different scientific domains, from metagenomics workflows producing hundreds of gigabytes of genome data to cultural heritage records with access restrictions for personally identifiable data. Specific *RO-Crate profiles* (section 2.2.6) may constrain serialization and publication expectations, and require additional contextual types and properties.

*2.2.3. Data Entities are described using Contextual Entities*

RO-Crate distinguishes between data and contextual entities[16] in a similar way to HTTP terminology's early attempt to separate *information* (data) and *non-information* (contextual) resources [31]. Data entities are usually files and directories located by relative IRI references within the RO-Crate Root, but they can also be Web resources or restricted data identified with absolute IRIs, including *Persistent Identifiers* (PIDs) [25].

As both types of entities are identified by IRIs, their distinction is allowed to be blurry; data entities can be located anywhere and be complex, while contextual entities can have a Web presence beyond their description inside the RO-Crate. For instance `https://orcid.org/0000-0002-1825-0097` is primarily an identifier for a person, but secondarily it is also a Web page and a way to refer to their academic work.

A particular IRI may appear as a contextual entity in one RO-Crate and as a data entity in another; the distinction lies in the fact that data entities can be considered to be *contained* or captured by that RO-Crate (*RO Content* in Figure 1), while contextual entities mainly *explain* an RO-Crate or its content (although this distinction is not a formal requirement).

---

[12]<https://www.researchobject.org/ro-crate/1.1/structure.html#ro-crate-metadata-file-ro-crate-metadatajson>
[13]<https://json-ld.org/#developers>
[14]<https://www.researchobject.org/ro-crate/1.1/root-data-entity.html#direct-properties-of-the-root-data-entity>
[15]<https://pages.github.com/>
[16]<https://www.researchobject.org/ro-crate/1.1/contextual-entities.html#contextual-vs-data-entities>

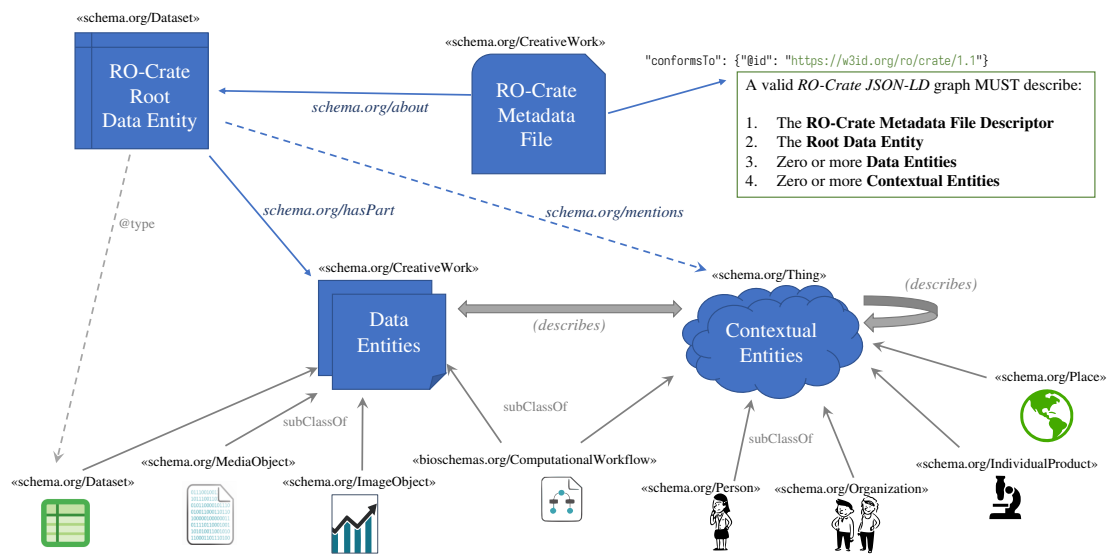

Fig. 2. **Simplified UML class diagram of RO-Crate.** The *RO-Crate Metadata File* conforms to a version of the specification; and contains a JSON-LD graph [32] that describes the entities that make up the RO-Crate. The *RO-Crate Root Data Entity* represent the Research Object as a dataset. The RO-Crate aggregates *data entities* (`hasPart`) which are further described using *contextual entities* (which may include aggregated and non-aggregated data entities). Multiple types and relations from Schema.org allow annotations to be more specific, including figures, nested datasets, computational workflows, people, organisations, instruments and places. Contextual entities not otherwise cross-referenced from other entities' properties (*describes*) can be grouped under the root entity (`mentions`).

In RO-Crate, a referenced contextual entity (e.g. a person identified by ORCID) should always be described within the RO-Crate Metadata File with at least a *type* and *name*, even where their PID might resolve to further Linked Data. This is so that clients are not required to follow every link for presentation purposes, for instance HTML rendering. Similarly any imported extension terms[17] would themselves also have a human-readable description in the case where their PID does not resolve to human-readable documentation.

Figure 2 shows a simplified UML class diagram of RO-Crate, highlighting the different types of data entities and contextual entities that can be aggregated and related. While an RO-Crate would usually contain one or more data entities (`hasPart`), it may also be a pure aggregation of contextual entities (`mentions`).

### 2.2.4. Guide through Recommended Practices

RO-Crate as a specification aims to build a set of recommended practices on how to practically apply existing standards in a common way to describe research outputs and their provenance, without having to learn each of the underlying technologies in detail.

As such, the RO-Crate 1.1[18] specification [33] can be seen as an opinionated and example-driven guide to writing Schema.org[19] [34] metadata as JSON-LD [32] (see section 2.3), which leaves it open for implementers to include additional metadata using other Schema.org types and properties, or even additional Linked Data vocabularies/ontologies or their own ad-hoc terms.

---

[17] <https://www.researchobject.org/ro-crate/1.1/appendix/jsonld.html#extending-ro-crate>
[18] <https://w3id.org/ro/crate/1.1>
[19] <https://schema.org/>

However the primary purpose of the RO-Crate specification is to assist developers in leveraging Linked Data principles for the focused purpose of describing Research Objects in a structured language, while reducing the steep learning curve otherwise associated in Semantic Web adaptation, like development of ontologies, identifiers, namespaces, and RDF serialization choices.

*2.2.5. Ensuring Simplicity*

One aim of RO-Crate is to be conceptually simple. This simplicity has been repeatedly checked and confirmed through an informal community review process. For instance, in the discussion on supporting ad-hoc vocabularies[20] in RO-Crate, the community explored potential Linked Data solutions. The conventional wisdom in RDF best practices[21] is to establish a vocabulary with a new URI namespace, formalised using RDF Schema[22] or OWL[23] ontologies. However, this may seem an excessive learning curve for non-experts in semantic knowledge representation, and the RO-Crate community instead agreed on a dual lightweight approach: (i) Document[24] how projects with their own Web-presence can make a pure HTML-based vocabulary, and (ii) provide a community-wide PID namespace under `https://w3id.org/ro/terms` that redirect to simple CSV files maintained in GitHub[25].

To further verify this idea of simplicity, we have formalised the RO-Crate definition (see *Appendix A)*). An important result of this exercise is that the underlying data structure of RO-Crate, although conceptually a graph, is represented as a depth-limited tree. This formalisation also emphasises the *boundedness* of the structure; namely, the fact that elements are specifically identified as being either semantically *contained* by the RO-Crate as *Data Entities* (`hasPart`) or mainly referenced (`mentions`) and typed as *external* to the Research Object as *Contextual Entities*. It is worth pointing out that this semantic containment can extend beyond the physical containment of files residing within the RO-Crate Root directory on a given storage system, as the RO-Crate data entities may include any data resource globally identifiable using IRIs.

*2.2.6. Extensibility and RO-Crate profiles*

The RO-Crate specification provides a core set of conventions to describe research outputs using types and properties applicable across scientific domains. However we have found that domain-specific use of RO-Crate will, implicitly or explicitly, form a specialized **profile** of RO-Crate; i.e., *a set of conventions, types and properties that are minimally required and one can expect to be present in that subset of RO-Crates.* For instance, RO-Crates used for exchange of workflows will have to contain a data entity of type `ComputationalWorkflow`, or cultural heritage records should have a `contentLocation`.

Making such profiles explicit allow further reliable programmatic consumption and generation of RO-Crates beyond the core types defined in the RO-Crate specification. Following the RO-Crate mantra of *guidance over strictness*, profiles are mainly *duck-typing* rather than strict syntactic or semantic types, but may also have corresponding machine-readable schemas at multiple levels (file formats, JSON, RDF shapes, RDFS/OWL semantics).

---

[20]<https://github.com/ResearchObject/ro-crate/issues/71>
[21]<https://www.w3.org/TR/swbp-vocab-pub/>
[22]<http://www.w3.org/TR/2014/REC-rdf-schema-20140225/>
[23]<http://www.w3.org/TR/2012/REC-owl2-overview-20121211/>
[24]<https://www.researchobject.org/ro-crate/1.1/appendix/jsonld.html#adding-new-or-ad-hoc-vocabulary-terms>
[25]<https://github.com/ResearchObject/ro-terms>

The next version of the RO-Crate specification 1.2 will define a formalization[26] for publishing and declaring conformance to RO-Crate profiles. Such a profile is primarily a human-readable document of before-mentioned expectations and conventions, but may also define a machine-readable profile as a **Profile Crate**: Another RO-Crate that describe the profile and in addition can list schemas for validation, compatible software, applicable repositories, serialization/packaging formats, extension vocabularies, custom JSON-LD contexts and examples (see for example the Workflow RO-Crate profile[27]).

In addition, there are sometimes existing domain-specific metadata formats, but they are either not RDF-based (and thus time-consuming to construct terms for in JSON-LD) or are at a different granularity level that might become overwhelming if represented directly in the RO-Crate Metadata file (e.g. W3C PROV bundle detailing every step execution of a workflow run [35]). RO-Crate allows such *alternative metadata files* to co-exist, and be described as data entities with references to the standards and vocabularies they conform to. This simplifies further programmatic consumption even where no filename or file extension conventions have emerged for those metadata formats.

Section 4 examines the observed specializations of RO-Crate use in several domains and their emerging profiles.

### 2.3. Technical implementation of the RO-Crate model

The RO-Crate conceptual model has been realised using JSON-LD and Schema.org in a prescriptive form as discussed in section 2.2. These technical choices were made to cater for simplicity from a developer perspective (as introduced in section 2.1).

JSON-LD[28] [32] provides a way to express Linked Data as a JSON structure, where a *context* provides mapping to RDF properties and classes. While JSON-LD cannot map arbitrary JSON structures to RDF, we found that it does lower the barrier compared to other RDF syntaxes, as the JSON syntax nowadays is a common and popular format for data exchange on the Web.

However, JSON-LD alone has too many degrees of freedom and hidden complexities for software developers to reliably produce and consume without specialised expertise or large RDF software frameworks. A large part of the RO-Crate specification is therefore dedicated to describing the acceptable subset of JSON structures.

#### 2.3.1. RO-Crate JSON-LD

RO-Crate mandates[29] the use of flattened, compacted JSON-LD in the RO-Crate Metadata file `ro-crate-metadata.json`[30] where a single `@graph` array contains all the data and contextual entities in a flat list. An example can be seen in the JSON-LD snippet in Listing 1 below, describing a simple RO-Crate containing data entities described using contextual entities:

---

[26]<https://www.researchobject.org/ro-crate/1.2-DRAFT/profiles>
[27]<https://w3id.org/workflowhub/workflow-ro-crate/>
[28]<https://json-ld.org/>
[29]<https://www.researchobject.org/ro-crate/1.1/appendix/jsonld.html>
[30]The avid reader may spot that the RO-Crate Metadata file use the extension `.json` instead of `.jsonld`, this is to emphasize the developer expectations as a JSON format, while the file's JSON-LD nature is secondary. See https://github.com/ResearchObject/ro-crate/issues/82

```
{ "@context": "https://w3id.org/ro/crate/1.1/context",
  "@graph": [
      { "@id": "ro-crate-metadata.json",
        "@type": "CreativeWork",
        "conformsTo": {"@id": "https://w3id.org/ro/crate/1.1"},
        "about": {"@id": "./"}
      },
      { "@id": "./",
        "@type": "Dataset",
        "name": "A simplified RO-Crate",
        "author": {"@id": "#alice"},
        "license": {"@id": "https://spdx.org/licenses/CC-BY-4.0"},
        "datePublished": "2021-11-02T16:04:43Z",
        "hasPart": [
          {"@id": "survey-responses-2019.csv"},
          {"@id": "https://example.com/pics/5707039334816454031_o.jpg"}
        ]
      },
      { "@id": "survey-responses-2019.csv",
        "@type": "File",
        "about": {"@id": "https://example.com/pics/5707039334816454031_o.jpg"},
        "author": {"@id": "#alice"}
      },
      { "@id": "https://example.com/pics/5707039334816454031_o.jpg",
        "@type": ["File", "ImageObject"],
        "contentLocation": {"@id": "http://sws.geonames.org/8152662/"},
        "author": {"@id": "https://orcid.org/0000-0002-1825-0097"}
      },
      { "@id": "#alice",
        "@type": "Person",
        "name": "Alice"
      },
      { "@id": "https://orcid.org/0000-0002-1825-0097",
        "@type": "Person",
        "name": "Josiah Carberry"
      },
      { "@id": "http://sws.geonames.org/8152662/",
        "@type": "Place",
        "name": "Catalina Park"
      },
      { "@id": "https://spdx.org/licenses/CC-BY-4.0",
        "@type": "CreativeWork",
        "name": "Creative Commons Attribution 4.0"
      }
  ]
}
```

**Listing 1**: Simplified[31] RO-Crate metadata file showing the flattened compacted JSON-LD `@graph` array containing the data entities and contextual entities, cross-referenced using `@id`. The `ro-crate-metadata.json` entity self-declares conformance with the RO-Crate specification using a versioned persistent identifier, further RO-Crate descriptions are on the root data entity `./` or any of the referenced data or contextual entities, as exemplified by the data entity `ImageObject` referencing contextual entities for `contentLocation` and `author` that differs from that of the overall RO-Crate. In this

---

[31] Recommended properties for types shown in Listing 1 also include `affiliation`, `citation`, `contactPoint`, `description`, `encodingFormat`, `funder`, `geo`, `identifier`, `keywords`, `publisher`; these properties and corresponding contextual entities are excluded here for brevity. See complete example https://www.researchobject.org/2021-packaging-research-artefacts-with-ro-crate/listing1/

example, `about` of the CSV data entity reference the `ImageObject`, which then take the roles of both a data entity and contextual entity. While `Person` entities ideally are identified with ORCID PIDs as for Josiah, `#alice` is here in contrast an RO-Crate local identifier, highlighting the pragmatic "just enough" Linked Data approach.

In this flattened profile of JSON-LD, each `{entity}` are directly under `@graph` and represents the RDF triples with a common *subject* (`@id`), mapped *properties* like `hasPart`, and *objects* — as either literal `"string"` values, referenced `{objects}` (which properties are listed in its own entity), or a JSON `[list]` of these. If processed as JSON-LD, this forms an RDF graph by matching the `@id` IRIs and applying the `@context` mapping to schema.org terms.

### 2.3.2. Flattened JSON-LD

When JSON-LD 1.0 [32] was proposed, one of the motivations was to seamlessly apply an RDF nature on top of regular JSON as frequently used by Web APIs. JSON objects in APIs are frequently nested with objects at multiple levels, and the perhaps most common form of JSON-LD is the compacted form[32] which follows this expectation (JSON-LD 1.1[33] further expands these capabilities, e.g. allowing nested `@context` definitions).

While this feature of JSON-LD can be seen as a way to "hide" its RDF nature, we found that the use of nested trees (e.g. a `Person` entity appearing as `author` of a `File` which nests under a `Dataset` with `hasPart`) counter-intuitively forces consumers to consider the JSON-LD as an RDF Graph, since an identified `Person` entity can appear at multiple and repeated points of the tree (e.g. author of multiple files), necessitating node merging or duplication, which can become complicated as this approach also invites the use of *blank nodes* (entities missing `@id`).

By comparison, a single flat `@graph` array approach, as required by RO-Crate, means that applications can choose to process and edit each entity as pure JSON by a simple lookup based on `@id`. At the same time, lifting all entities to the same level reflects the Research Object principles [11] in that describing the context and provenance is just as important as describing the data, and the requirement of `@id` of every entity forces RO-Crate generators to consciously consider existing IRIs and identifiers[34].

### 2.3.3. JSON-LD context

In JSON-LD, the `@context` is a reference to another JSON-LD document that provides mapping from JSON keys to Linked Data term IRIs, and can enable various JSON-LD directives to cater for customized JSON structures for translating to RDF.

RO-Crate reuses vocabulary terms and IRIs from Schema.org, but provides its own versioned JSON-LD context[35], which has a flat list with the mapping from JSON-LD keys to their URI equivalents (e.g. key `"author"` maps to the http://schema.org/author property).

The rationale behind this decision is to support JSON-based RO-Crate applications that are largely unaware of JSON-LD, that still may want to process the `@context` to find or add Linked Data definitions of otherwise unknown properties and types. Not reusing the official Schema.org context means RO-Crate is also able to map in additional vocabularies where needed, namely the *Portland Common Data Model* (PCDM) [36] for repositories and Bioschemas [37] for describing computational workflows. RO-Crate profiles may extend[36] the `@context` to re-use additional domain-specific ontologies.

---

[32]<https://json-ld.org/spec/REC/json-ld/20140116/#compacted-document-form>
[33]<https://www.w3.org/TR/2020/REC-json-ld11-20200716/>
[34]<https://www.researchobject.org/ro-crate/1.1/appendix/jsonld.html#describing-entities-in-json-ld>
[35]<https://w3id.org/ro/crate/1.1/context>

Similarly, while the Schema.org context currently[37] have `"@type": "@id"` annotations for object properties, RO-Crate JSON-LD distinguishes explicitly between references to other entities (`{"@id": "#alice"}`) and string values (`"Alice"`) — meaning RO-Crate applications can find references for corresponding entities and IRIs without parsing the `@context` to understand a particular property. Notably this is exploited by the `ro-crate-html-js` [38] tool to provide reliable HTML rendering for otherwise unknown properties and types.

*2.4. RO-Crate Community*

The RO-Crate conceptual model, implementation and best practices are developed by a growing community of researchers, developers and publishers. RO-Crate's community is a key aspect of its effectiveness in making research artefacts FAIR. Fundamentally, the community provides the overall context of the implementation and model and ensures its interoperability.

The RO-Crate community consists of:
1. a diverse set of people representing a variety of stakeholders;
2. a set of collective norms;
3. an open platform that facilitates communication (GitHub, Google Docs, monthly teleconferences).

*2.4.1. People*

The initial concept of RO-Crate was formed at the first Workshop on Research Objects (RO2018[38]), held as part of the IEEE conference on eScience. This workshop followed up on considerations made at a Research Data Alliance (RDA) meeting on Research Data Packaging[39] that found similar goals across multiple data packaging efforts [13]: simplicity, structured metadata and the use of JSON-LD.

An important outcome of discussions that took place at RO2018 was the conclusion that the original Wf4Ever Research Object ontologies [12], in principle sufficient for packaging research artefacts with rich descriptions, were, in practice, considered inaccessible for regular programmers (e.g., Web developers) and in danger of being incomprehensible for domain scientists due to their reliance on Semantic Web technologies and other ontologies.

DataCrate [39] was presented at RO2018 as a promising lightweight alternative approach, and an agreement was made by a group of volunteers to attempt building what was initially called *"RO Lite"* as a combination of DataCrate's implementation and Research Object's principles.

This group, originally made up of library and Semantic Web experts, has subsequently grown to include domain scientists, developers, publishers and more. This perspective of multiple views led to the specification being used in a variety of domains, from bioinformatics and regulatory submissions to humanities and cultural heritage preservation.

The RO-Crate community is strongly engaged with the European-wide biology/bioinformatics collaborative e-Infrastructure, ELIXIR [40], along with European Open Science Cloud (EOSC[40]) projects including EOSC-Life[41], FAIRplus[42], CS3MESH4EOSC[43] and BY-COVID[44]. RO-Crate has also established collaborations with Bioschemas [37], GA4GH [41], OpenAIRE [42] and multiple H2020 projects.

---

[36]<https://www.researchobject.org/ro-crate/1.1/appendix/jsonld.html#extending-ro-crate>
[37]<https://schema.org/version/13.0/schemaorg-current-http.jsonld>
[38]<https://www.researchobject.org/ro2018/>
[39]<https://rd-alliance.org/approaches-research-data-packaging-rda-11th-plenary-bof-meeting>
[40]<https://eosc.eu/>

A key set of stakeholders are developers: the RO-Crate community has made a point of attracting developers who can implement the specifications but, importantly, keeps "developer user experience" in mind. This means that the specifications are straightforward to implement and thus do not require expertise in technologies that are not widely deployed.

This notion of catering to "developer user experience" is an example of the set of norms that have developed and now define the community.

*2.4.2. Norms*

The RO-Crate community is driven by informal conventions and notions that are prevalent but not neccessarily written down. Here, we distil what we as authors believe are the critical set of norms that have facilitated the development of RO-Crate and contributed to the ability for RO-Crate research packages to be FAIR. This is not to say that there are no other norms within the community nor that everyone in the community holds these uniformly. Instead, what we emphasise is that these norms are helpful and also shaped by community practices.

1. Simplicity
2. Developer friendliness
3. Focus on examples and best practices rather than rigorous specification
4. Reuse "just enough" Web standards

A core norm of RO-Crate is that of **simplicity**, which sets the scene for how we guide developers to structure metadata with RO-Crate. We focus mainly on documenting simple approaches to the most common use cases, such as authors having an affiliation. This norm also influences our take on **developer friendliness**; for instance, we are using the Web-native JSON format, allowing only a few of JSON-LD's flexible Linked Data features. Moreover, the RO-Crate documentation is largely built up by **examples** showcasing **best practices**, rather than rigorous specifications. We build on existing **Web standards** that themselves are defined rigorously, which we utilise ***"just enough"*** in order to benefit from the advantages of Linked Data (e.g., extensions by namespaced vocabularies), without imposing too many developer choices or uncertainties (e.g., having to choose between the many RDF syntaxes).

While the above norms alone could easily lead to the creation of "yet another" JSON format, we keep the goal of **FAIR interoperability** of the captured metadata, and therefore follow closely FAIR best practices and current developments such as data citations, PIDs, open repositories and recommendations for sharing research outputs and software.

*2.4.3. Open Platforms*

The critical infrastructure that enables the community around RO-Crate is the use of open development platforms. This underpins the importance of open community access to supporting FAIR. Specifically, it is difficult to build and consume FAIR research artefacts without being able to access the specifications, understand how they are developed, know about any potential implementation issues, and discuss usage to evolve best practices.

---

[41]<https://www.eosc-life.eu/>
[42]<https://fairplus-project.eu/>
[43]<https://cs3mesh4eosc.eu/>
[44]<https://by-covid.eu/>

The development of RO-Crate was driven by capturing documentation of real-life examples and best practices rather than creating a rigorous specification. At the same time, we agreed to be opinionated on the syntactic form to reduce the jungle of implementation choices; we wanted to keep the important aspects of Linked Data to adhere to the FAIR principles while retaining the option of combining and extending the structured metadata using the existing Semantic Web stack, not just build a standalone JSON format.

Further work during 2019 started adapting the DataCrate documentation through a more collaborative and exploratory *RO Lite* phase, initially using Google Docs for review and discussion, then moving to GitHub as a collaboration space for developing what is now the RO-Crate specification, maintained as Markdown[45] in GitHub Pages and published through Zenodo.

In addition to the typical Open Source-style development with GitHub issues and pull requests, the RO-Crate Community have, at time of writing, two regular monthly calls, a Slack channel and a mailing list for coordinating the project; also many of its participants collaborate on RO-Crate at multiple conferences and coding events such as the ELIXIR BioHackathon[46]. The community is jointly developing the RO-Crate specification and Open Source tools, as well as providing support and considering new use cases. The RO-Crate Community[47] is open for anyone to join, to equally participate under a code of conduct, and as of October 2021 has more than 50 members (see Appendix B).

## 3. RO-Crate Tooling

The work of the community has led to the development of a number of tools for creating and using RO-Crates. Table 1 shows the current set of implementations. Reviewing this list, one can see that tools support commonly used programming languages, including Python, JavaScript, and Ruby. Additionally, these tools can be integrated into commonly used research environments, in particular, the command line tool *ro-crate-html-js* [38] for creating a human-readable preview of an RO-Crate as a sidecar HTML file. Furthermore, there are tools that cater to end-users (*Describo* [22], *WorkflowHub* [48]), in order to simplify creating and managing RO-Crate. For example, Describo was developed to help researchers of the Australian Criminal Characters project[48] annotate historical prisoner records to gain greater insight into the history of Australia [61].

While the development of these tools is promising, our analysis of their maturity status shows that the majority of them are in the Beta stage. This is partly due to the fact that the RO-Crate specification itself only recently reached 1.0 status, in November 2019 [62]. Now that there is a fixed point of reference: With version 1.1 (October 2020) [63] RO-Crate has stabilised based on feedback from application development, and now we are seeing a further increase in the maturity of these tools, along with the creation of new ones.

Given the stage of the specification, these tools have been primarily targeting developers, essentially providing them with the core libraries for working with RO-Crate. Another target has been that of research data managers who need to manage and curate large amounts of data.

---

[45]<https://github.com/researchobject/ro-crate/>
[46]<https://biohackathon-europe.org/>
[47]<https://www.researchobject.org/ro-crate/community>
[48]<https://criminalcharacters.com/>

| Tool name<br>*Brief Description* | Targets | Language / Platform | Status |
|---|---|---|---|
| **Describo** [22] | Research Data Managers | NodeJS (Desktop) | RC |
| *Interactive desktop application to create, update and export RO-Crates for different profiles* | | | |
| **Describo Online** [43] | Platform developers | NodeJS (Web) | Alpha |
| *Web-based application to create RO-Crates using cloud storage* | | | |
| **ro-crate-excel** [44] | Data managers | JavaScript | |
| *Command-line tool to create/edit RO-Crates with spreadsheets* | | | |
| **ro-crate-html-js** [38] | Developers | JavaScript | Beta |
| *HTML rendering of RO-Crate* | | | |
| **ro-crate-js** [45] | Research Data Managers | JavaScript | Alpha |
| *Library for creating/manipulating crates; basic validation code* | | | |
| **ro-crate-ruby** [46] | Developers | Ruby | Beta |
| *Ruby library for reading/writing RO-Crate, with workflow support* | | | |
| **ro-crate-py** [47] | Developers | Python | Beta |
| *Object-oriented Python library for reading/writing RO-Crate and use by Jupyter Notebook* | | | |
| **WorkflowHub** [48] | Workflow users | Ruby | Beta |
| *Workflow repository; imports and exports Workflow RO-Crate* | | | |
| **Life Monitor** [49] | Workflow developers | Python | Alpha |
| *Workflow testing and monitoring service; Workflow Testing profile of RO-Crate* | | | |
| **SCHeMa** [50] | Workflow users | PHP | Alpha |
| *Workflow execution using RO-Crate as exchange mechanism* | | | |
| **galaxy2cwl** [51] | Workflow developers | Python | Alpha |
| *Wraps Galaxy workflow as Workflow RO-Crate* | | | |
| **Modern PARADISEC** [52] | Repository managers | Platform | Beta |
| *Cultural Heritage portal based on OCFL and RO-Crate* | | | |
| **ONI express** [53] | Repository managers | Platform | Beta |
| *Platform for publishing data and documents stored in an OCFL repository via a Web interface* | | | |
| **ocfl-tools** [54] | Developers | JavaScript (CLI) | Beta |
| *Tools for managing RO-Crates in an OCFL repository* | | | |
| **RO Composer** [55] | Repository developers | Java | Alpha |
| *REST API for gradually building ROs for given profile* | | | |
| **RDA maDMP Mapper** [56] | Data Management Plan users | Python | Beta |
| *Mapping between machine-actionable data management plans (maDMP) and RO-Crate* [57] | | | |
| **Ro-Crate_2_ma-DMP** [58] | Data Management Plan users | Python | Beta |
| *Convert between machine-actionable data management plans (maDMP) and RO-Crate* | | | |
| **CheckMyCrate** [59] | Developers | Python (CLI) | Alpha |
| *Validation according to Workflow RO-Crate profile* | | | |
| **RO-Crates-and-Excel** [60] | Data Managers | Java (CLI) | Alpha |
| *Describe column/data details of spreadsheets as RO-Crate using DataCube vocabulary* | | | |

Table 1

Applications and libraries implementing RO-Crate, targeting different types of users across multiple programming languages. Status is indicative as assessed by this work (Alpha < Beta < Release Candidate (RC) < Release).

## 4. Profiles of RO-Crate in use

RO-Crate fundamentally forms part of an infrastructure to help build FAIR research artefacts. In other words, the key question is whether RO-Crate can be used to share and (re)use research artefacts. Here we look at three research domains where RO-Crate is being applied: Bioinformatics, Regulatory Science and Cultural Heritage. In addition, we note how RO-Crate may have an important role as part of machine-actionable data management plans and institutional repositories.

From these varied uses of RO-Crate we observe natural differences in their detail level and the type of entities described by the RO-Crate. For instance, on submission of an RO-Crate to a workflow repository, it is reasonable to expect the RO-Crate to contain at least one workflow, ideally with a declared licence and workflow language. Specific additional recommendations such as on identifiers is also needed to meet the emerging requirements of FAIR Digital Objects[49]. Work has now begun[50] to formalise these different *profiles* of RO-Crates, which may impose additional constraints based on the needs of a specific domain or use case.

*4.1. Bioinformatics workflows*

WorkflowHub.eu[51] is a European cross-domain registry of computational workflows, supported by European Open Science Cloud projects, e.g. EOSC-Life[52], and research infrastructures including the pan-European bioinformatics network ELIXIR[53] [40]. As part of promoting workflows as reusable tools, WorkflowHub includes documentation and high-level rendering of the workflow structure independent of its native workflow definition format. The rationale is that a domain scientist can browse all relevant workflows for their domain, before narrowing down their workflow engine requirements. As such, the WorkflowHub is intended largely as a registry of workflows already deposited in repositories specific to particular workflow languages and domains, such as UseGalaxy.eu [64] and Nextflow nf-core [65].

We here describe three different RO-Crate profiles developed for use with WorkflowHub.

*4.1.1. Profile for describing workflows*

Being cross-domain, WorkflowHub has to cater for many different workflow systems. Many of these, for instance Nextflow [66] and Snakemake [67], by virtue of their script-like nature, reference multiple neighbouring files typically maintained in a GitHub repository. This calls for a data exchange method that allows keeping related files together. WorkflowHub has tackled this problem by adopting RO-Crate as the packaging mechanism [68], typing and annotating the constituent files of a workflow and — crucially — marking up the workflow language, as many workflow engines use common file extensions like *.xml and *.json. Workflows are further described with authors, license, diagram previews and a listing of their inputs and outputs. RO-Crates can thus be used for interoperable deposition of workflows to WorkflowHub, but are also used as an archive for downloading workflows, embedding metadata registered with the WorkflowHub entry and translated workflow files such as abstract Common Workflow Language (CWL) [69] definitions and diagrams [70].

---

[49]<https://fairdo.org/>
[50]<https://github.com/ResearchObject/ro-crate/issues/153>
[51]<https://workflowhub.eu/>
[52]<https://www.eosc-life.eu/>
[53]<https://elixir-europe.org/>

RO-Crate acts therefore as an interoperability layer between registries, repositories and users in WorkflowHub. The iterative development between WorkflowHub developers and the RO-Crate community heavily informed the creation of the Bioschemas [37] profile for Computational Workflows[54], which again informed the RO-Crate 1.1 specification on workflows[55] and led to the RO-Crate Python library [47] and WorkflowHub's **Workflow RO-Crate profile**[56], which, in a similar fashion to RO-Crate itself, recommends which workflow resources and descriptions are required. This co-development across project boundaries exemplifies the drive for simplicity and for establishing best practices.

*4.1.2. Profile for recording workflow runs*

RO-Crates in WorkflowHub have so far been focused on workflows that are ready to be run, and development of WorkflowHub is now creating a **Workflow Run RO-Crate profile** for the purposes of benchmarking, testing and executing workflows. As such, RO-Crate serves as a container of both a *workflow definition* that may be executed and of a particular *workflow execution with test results*.

This workflow run profile is a continuation of our previous work with capturing workflow provenance in a Research Object in CWLProv [35] and TavernaPROV [71]. In both cases, we used the PROV Ontology [59], including details of every task execution with all the intermediate data, which required significant workflow engine integration.[57]

Simplifying from the CWLProv approach, the planned Workflow Run RO-Crate profile will use a high level Schema.org provenance[58] for the input/output boundary of the overall workflow execution. This *Level 1 workflow provenance* [35] can be expressed generally across workflow languages with minimal workflow engine changes, with the option of more detailed provenance traces as separate PROV artefacts in the RO-Crate as data entities. In the current development of Specimen Data Refinery[59] [73] these RO-Crates will document the text recognition workflow runs of digitised biological specimens, exposed as FAIR Digital Objects [74].

WorkflowHub has recently enabled minting of Digital Object Identifiers (DOIs), a PID commonly used for scholarly artefacts, for registered workflows, e.g. `10.48546/workflowhub.workflow.56.1` [75], lowering the barrier for citing workflows as computational methods along with their FAIR metadata – captured within an RO-Crate. While it is not an aim for WorkflowHub to be a repository of workflow runs and their data, RO-Crates of *exemplar workflow runs* serve as useful workflow documentation, as well as being an exchange mechanism that preserves FAIR metadata in a diverse workflow execution environment.

*4.1.3. Profile for testing workflows*

The value of computational workflows, however, is potentially undermined by the "collapse" over time of the software and services they depend upon: for instance, software dependencies can change in a non-backwards-compatible manner, or active maintenance may cease; an external resource, such as a reference index or a database query service, could shift to a different URL or modify its access protocol; or the workflow itself may develop hard-to-find bugs as it is updated. This *workflow*

---

[54]<https://bioschemas.org/profiles/ComputationalWorkflow/1.0-RELEASE/>

[55]<https://www.researchobject.org/ro-crate/1.1/workflows.html>

[56]<https://w3id.org/workflowhub/workflow-ro-crate/1.0>

[57]CWLProv and TavernaProv predate RO-Crate, but use RO-Bundle[72], a similar Research Object packaging method with JSON-LD metadata.

[58]<https://www.researchobject.org/ro-crate/1.1/provenance.html#software-used-to-create-files>

[59]<https://github.com/DiSSCo/SDR>

*decay* can take a big toll on the workflow's reusability and on the reproducibility of any processes it evokes [76].

For this reason, WorkflowHub is complemented by a monitoring and testing service called LifeMonitor[49], also supported by EOSC-Life. LifeMonitor's main goal is to assist in the creation, periodic execution and monitoring of workflow tests, enabling the early detection of software collapse in order to minimise its detrimental effects. The communication of metadata related to workflow testing is achieved through the adoption of a **Workflow Testing RO-Crate profile**[60] stacked on top of the *Workflow RO-Crate* profile. This further specialisation of Workflow RO-Crate allows to specify additional testing-related entities (test suites, instances, services, etc.), leveraging RO-Crate's extension mechanism[61] through the addition of terms from custom namespaces.

In addition to showcasing RO-Crate's extensibility, the testing profile is an example of the format's flexibility and adaptability to the different needs of the research community. Though ultimately related to a computational workflow, in fact, most of the testing-specific entities are more about describing a protocol for interacting with a monitoring service than a set of research outputs and its associated metadata. Indeed, one of LifeMonitor's main functionalities is monitoring and reporting on test suites running on existing Continuous Integration (CI) services, which is described in terms of service URLs and job identifiers in the testing profile. In principle, in this context, data could disappear altogether, leading to an RO-Crate consisting entirely of contextual entities. Such an RO-Crate acts more as an exchange format for communication between services (WorkflowHub and LifeMonitor) than as an aggregator for research data and metadata, providing a good example of the format's high versatility.

*4.2. Regulatory Sciences*

BioCompute Objects[62] (BCO) [77] is a community-led effort to standardise submissions of computational workflows to biomedical regulators. For instance, a genomics sequencing pipeline, as part of a personalised cancer treatment study, can be submitted to the US Food and Drugs Administration (FDA) for approval. BCOs are formalised in the standard IEEE 2791-2020 [78] as a combination of JSON Schemas[63] that define the structure of JSON metadata files describing exemplar workflow runs in detail, covering aspects such as the usability and error domain of the workflow, its runtime requirements, the reference datasets used and representative output data produced.

BCOs provide a structured view over a particular workflow, informing regulators about its workings independently of the underlying workflow definition language. However, BCOs have only limited support for additional metadata.[64] For instance, while the BCO itself can indicate authors and contributors, and in particular regulators and their review decisions, it cannot describe the provenance of individual data files or workflow definitions.

As a custom JSON format, BCOs cannot be extended with Linked Data concepts, except by adding an additional top-level JSON object formalised in another JSON Schema. A BCO and workflow submitted by upload to a regulator will also frequently consist of multiple cross-related files. Crucially, there is no way to tell whether a given *.json file is a BCO file, except by reading its content and check for its spec_version.

---

[60]<https://lifemonitor.eu/workflow_testing_ro_crate>
[61]<https://www.researchobject.org/ro-crate/1.1/appendix/jsonld.html#extending-ro-crate>
[62]<https://biocomputeobject.org/>
[63]<https://opensource.ieee.org/2791-object/ieee-2791-schema/>
[64]IEEE 2791-2020 do permit user extensions in the *extension domain* by referencing additional JSON Schemas.

We can then consider how a BCO and its referenced artefacts can be packaged and transferred following FAIR principles. **BCO RO-Crate**[65][79], part of the BioCompute Object user guides, defines a set of best practices for wrapping a BCO with a workflow, together with its exemplar outputs in an RO-Crate, which then provides typing and additional provenance metadata of the individual files, workflow definition, referenced data and the BCO metadata itself.

Here the BCO is responsible for describing the *purpose* of a workflow and its run at an abstraction level suitable for a domain scientist, while the more open-ended RO-Crate describes the surroundings of the workflow, classifying and relating its resources and providing provenance of their existence beyond the BCO. This emerging *separation of concerns* is shown in Figure 3, and highlights how RO-Crate is used side-by-side of existing standards and tooling, even where there are apparent partial overlaps.

A similar separation of concerns can be found if considering the RO-Crate as a set of files, where the *transport-level* metadata, such as checksum of files, are delegated to separate BagIt[66] manifests, a standard focusing on the preservation challenges of digital libraries [26]. As such, RO-Crate metadata files are not required to iterate all the files in their folder hierarchy, only those that benefit from being described.

Specifically, a BCO description alone is insufficient for reliable re-execution of a workflow, which would need a compatible workflow engine depending on the original workflow definition language, so IEEE 2791 recommends using Common Workflow Language (CWL) [69] for interoperable pipeline execution. CWL itself relies on tool packaging in software containers using Docker[67] or Conda[68]. Thus, we can consider BCO RO-Crate as a stack: transport-level manifests of files (BagIt), provenance, typing and context of those files (RO-Crate), workflow overview and purpose (BCO), interoperable workflow definition (CWL) and tool distribution (Docker).

### 4.3. Digital Humanities: Cultural Heritage

The Pacific And Regional Archive for Digital Sources in Endangered Cultures (PARADISEC[69]) [82] maintains a repository of more than 500,000 files documenting endangered languages across more than 16,000 items, collected and digitized over many years by researchers interviewing and recording native speakers across the region.

The Modern PARADISEC demonstrator[70] has been proposed[71] as an update to the 18 year old infrastructure, to also help long-term preservation of these artefacts in their digital form. The demonstrator uses RO-Crate to describe the overall structure and to capture the metadata of each item. The existing PARADISEC data collection has been ported and captured as RO-Crates. A Web portal then exposes the repository and its entries by indexing the RO-Crate metadata files, presenting a domain-specific view of the items — the RO-Crate is "hidden" and does not change the user interface.

---

[65]<https://biocompute-objects.github.io/bco-ro-crate/>
[66]<https://www.researchobject.org/ro-crate/1.1/appendix/implementation-notes.html#adding-ro-crate-to-bagit>
[67]<https://www.docker.com/>
[68]<https://docs.conda.io/>
[69]<https://www.paradisec.org.au/>
[70]<https://mod.paradisec.org.au/>
[71]<https://arkisto-platform.github.io/case-studies/paradisec/>

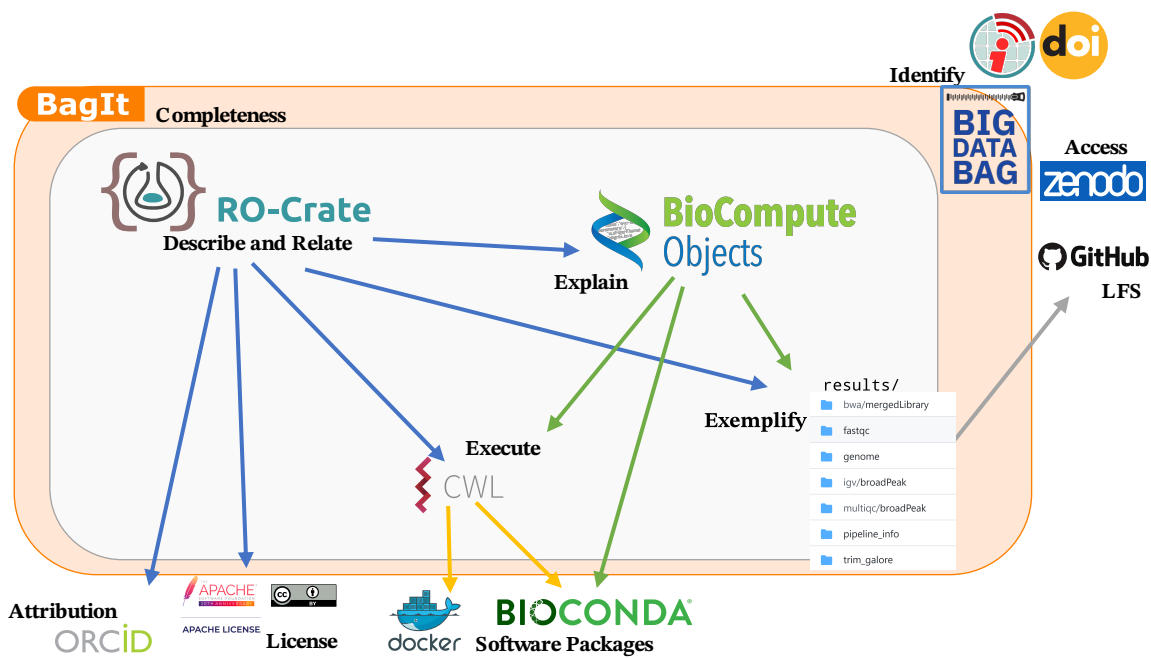

Fig. 3. **Separation of Concerns in BCO RO-Crate**. BioCompute Object (IEEE2791) is a JSON file that structurally explains the purpose and implementation of a computational workflow, for instance implemented in Common Workflow Language (CWL), that installs the workflow's software dependencies as Docker containers or BioConda packages. An example execution of the workflow shows the different kinds of result outputs, which may be external, using GitHub LFS [80] to support larger data. RO-Crate gathers all these local and external resources, relating them and giving individual descriptions, for instance permanent DOI identifiers for reused datasets accessed from Zenodo, but also adding external identifiers to attribute authors using ORCID or to identify which licences apply to individual resources. The RO-Crate and its local files are captured in a BagIt whose checksum ensures completeness, combined with Big Data Bag [81] features to "complete" the bag with large external files such as the workflow outputs.

The PARADISEC use case takes advantage of several RO-Crate features and principles. Firstly, the transcribed metadata are now independent of the PARADISEC platform and can be archived, preserved and processed in its own right, using Schema.org as base vocabulary and extended with PARADISEC-specific terms.

In this approach, RO-Crate is the holder of itemised metadata, stored in regular files that are organised using Oxford Common File Layout[72] (OCFL) [27], which ensures file integrity and versioning on a regular shared file system. This lightweight infrastructure also gives flexibility for future developments and maintenance. For example a consumer can use Linked Data software such as a graph database and query the whole corpora using SPARQL triple patterns across multiple RO-Crates. For long term digital preservation, beyond the lifetime of PARADISEC portals, a "last resort" fallback is storing the generic RO-Crate HTML preview [38]. Such human-readable rendering of RO-Crates can be hosted as static files by any Web server, in line with the approach taken by the Endings Project.[73]

---

[72]<https://ocfl.io/1.0/spec/>

[73]The Endings Project <https://endings.uvic.ca/> is a five-year project funded by the Social Sciences and Humanities Research Council (SSHRC) that is creating tools, principles, policies and recommendations for digital scholarship practitioners to create accessible, stable, long-lasting resources in the humanities.

*4.4. Machine-actionable Data Management Plans*

Machine-actionable Data Management Plans (maDMPs) have been proposed as an improvement to automate FAIR data management tasks in research [83]; maDMPs use PIDs and controlled vocabularies to describe what happens to data over the research life cycle [84]. The Research Data Alliance's *DMP Common Standard* for maDMPs [85] is one such formalisation for expressing maDMPs, which can be expressed as Linked Data using the DMP Common Standard Ontology [86], a specialisation of the W3C Data Catalog Vocabulary (DCAT) [87]. RDA maDMPs are usually expressed using regular JSON, conforming to the DMP JSON Schema.

A mapping has been produced between Research Object Crates and Machine-actionable Data Management Plans [57], implemented by the RO-Crate RDA maDMP Mapper [56]. A similar mapping has been implemented by `RO-Crate_2_ma-DMP` [58]. In both cases, a maDMP can be converted to a RO-Crate, or vice versa. In [57] this functionality caters for two use cases:

1. Start a skeleton data management plan based on an existing RO-Crate dataset, e.g. from an RO-Crate from WorkflowHub.
2. Instantiate an RO-Crate based on a data management plan.

An important nuance here is that data management plans are (ideally) written in *advance* of data production, while RO-Crates are typically created to describe data *after* it has been generated. What is significant to note in this approach is the importance of **templating** in order to make both tasks automatable and achievable, and how RO-Crate can fit into earlier stages of the research life cycle.

*4.5. Institutional data repositories – Harvard Data Commons*

The concept of a **Data Commons** for research collaboration was originally defined as "*cyber-infrastructure that co-locates data, storage, and computing infrastructure with commonly used tools for analysing and sharing data to create an interoperable resource for the research community*" [88]. More recently, Data Commons has been established to mean integration of active data-intensive research with data management and archival best practices, along with a supporting computational infrastructure. Furthermore, the Commons features tools and services, such as computation clusters and storage for scalability, data repositories for disseminating and preserving regular, but also large or sensitive datasets, and other research assets. Multiple initiatives were undertaken to create Data Commons on national, research, and institutional levels. For example, the Australian Research Data Commons (ARDC)[74] [89] is a national initiative that enables local researchers and industries to access computing infrastructure, training, and curated datasets for data-intensive research. NCI's Genomic Data Commons[75] (GDC) [90] provides the cancer research community with access to a vast volume of genomic and clinical data. Initiatives such as Research Data Alliance (RDA) Global Open Research Commons[76] propose standards for the implementation of Data Commons to prevent them becoming "data silos" and thus, enable interoperability from one Data Commons to another.

**Harvard Data Commons** [91] aims to address the challenges of data access and cross-disciplinary research within a research institution. It brings together multiple institutional schools, libraries, computing centres and the Harvard Dataverse[77] data repository. Dataverse[78] [92] is a free and open-source software platform to archive, share and cite research data. The Harvard Dataverse repository is the largest of 70 Dataverse installations worldwide, containing over 120K datasets with about 1.3M data files (as of 2021-11-16). Working toward the goal of facilitating collaboration and data discoverability and management within the university, Harvard Data Commons has the following primary objectives:

1. the integration of Harvard Research Computing with Harvard Dataverse by leveraging Globus endpoints [93]; this will allow an automatic transfer of large datasets to the repository. In some cases, only the metadata will be transferred while the data stays stored in remote storage;
2. support for advanced research workflows and providing packaging options for assets such as code and workflows in the Harvard Dataverse repository to enable reproducibility and reuse, and
3. interation of repositories supported by Harvard, which include DASH[79], the open access institutional repository, the Digital Repository Services (DRS) for preserving digital asset collections, and the Harvard Dataverse.

---

[74]<https://ardc.edu.au>
[75]<https://gdc.cancer.gov/>
[76]<https://www.rd-alliance.org/groups/global-open-research-commons-ig>
[77]<https://dataverse.harvard.edu/>
[78]<https://dataverse.org/>

Particularly relevant to this article is the second objective of the Harvard Data Commons, which aims to support the deposit of research artefacts to Harvard Dataverse with sufficient information in the metadata to allow their future reuse (Figure 4). To support the incorporation of data, code, and other artefacts from various institutional infrastructures, Harvard Data Commons is currently working on RO-Crate adaptation. The RO-Crate metadata provides the necessary structure to make all research artefacts FAIR. The Dataverse software already has extensive support for metadata, including the Data Documentation Initiative (DDI), Dublin Core, DataCite, and Schema.org. Incorporating RO-Crate, which has the flexibility to describe a wide range of research resources, will facilitate their seamless transition from one infrastructure to the other within the Harvard Data Commons.

Even though the Harvard Data Commons is specific to Harvard University, the overall vision and the three objectives can be abstracted and applied to other universities or research organisations. The Commons will be designed and implemented using standards and commonly-used approaches to make it interoperable and reusable by others.

## 5. Related Work

With the increasing digitisation of research processes, there has been a significant call for the wider adoption of interoperable sharing of data and its associated metadata. We refer to [94] for a comprehensive overview and recommendations, in particular for data; notably that review highlights the wide variety of metadata and documentation that the literature prescribes for enabling data reuse. Likewise, we suggest [95] that covers the importance of metadata standards in reproducible computational research.

Here we focus on approaches for bundling research artefacts along with their metadata. This notion of publishing compound objects for scholarly communication has a long history behind it [96] [97], but recent approaches have followed three main strands: 1) publishing to centralised repositories; 2) packaging approaches similar to RO-Crate; and 3) bundling the computational workflow around a scientific experiment.

---

[79]<https://dash.harvard.edu/>

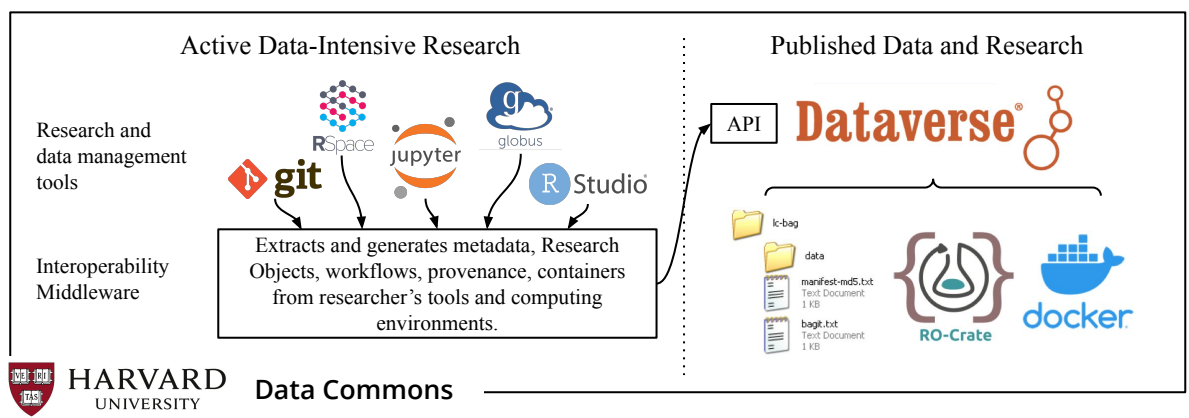

Fig. 4. **One aspect of Harvard Data Commons**. Automatic encapsulation and deposit of artefacts from data management tools used during active research at the Harvard Dataverse repository.

*5.1. Bundling and Packaging Digital Research Artefacts*

Early work making the case for publishing compound scholarly communication units [97] led to the development of the Object Re-Use and Exchange model[80] (OAI-ORE), providing a structured **resource map** of the digital artefacts that together support a scholarly output.

The challenge of describing computational workflows was one of the main motivations for the early proposal of *Research Objects* (RO) [11] as first-class citizens for sharing and publishing. The RO approach involves bundling datasets, workflows, scripts and results along with traditional dissemination materials like journal articles and presentations, forming a single package. Crucially, these resources are not just gathered, but also individually typed, described and related to each other using semantic vocabularies. As pointed out in [11] an open-ended *Linked Data* approach is not sufficient for scholarly communication: a common data model is also needed in addition to common and best practices for managing and annotating lifecycle, ownership, versioning and attributions.

Considering the FAIR principles [5], we can say with hindsight that the initial RO approaches strongly targeted *Interoperability*, with a particular focus on the reproducibility of *in-silico experiments* involving computational workflows and the reuse of existing RDF vocabularies.

The first implementation of Research Objects for sharing workflows in myExperiment [98] was based on RDF ontologies [99], building on Dublin Core, FOAF, SIOC, Creative Commons and OAI-ORE to form myExperiment ontologies for describing social networking, attribution and credit, annotations, aggregation packs, experiments, view statistics, contributions, and workflow components [100].

This initially workflow-centric approach was further formalized as the Wf4Ever Research Object Model [12], which is a general-purpose research artefact description framework. This model is based on existing ontologies (FOAF, Dublin Core Terms, OAI-ORE and AO/OAC precursors to the W3C Web Annotation Model [101]) and adds specializations for workflow models and executions using W3C PROV-O [102]. The Research Object statements are saved in a *manifest* (the OAI-ORE *resource map*), with additional annotation resources containing user-provided details such as title and description.

---

[80]<http://www.openarchives.org/ore/1.0/primer>

We now claim that one barrier for wider adoption of the Wf4Eer Research Object model for general packaging digital research artefacts was exactly this re-use of multiple existing vocabularies (FAIR principle I2: *Metadata use vocabularies that follow FAIR principles*), which in itself is recogized as a challenge [103]. Adapters of the Wf4Ever RO model would have to navigate documentation of multiple overlapping ontologies, in addition to facing the usual Semantic Web development choices for RDF serialization formats, identifier minting and publishing resources on the Web.

Several developments for Research Objects improved on this situation, such as ROHub used by Earth Sciences [104], which provides a interactive user-interface for making Research Objects, along with Research Object Bundle [72] (RO Bundle), which is a ZIP-archive embedding data files and a JSON-LD serialization of the manifest with mappings for a limited set of terms. RO Bundle was also used for storing detailed workflow run provenance (TavernaPROV [71]).

RO-Bundle evolved to Research Object BagIt archives,[81] a variant of RO Bundle as a BagIt archive [26], used by Big Data Bags [81], CWLProv [35] and WholeTale [105] [106].

### 5.2. FAIR Digital Objects

FAIR Digital Objects (FDO) [74] have been proposed as a conceptual framework for making digital resources available in a Digital Objects (DO) architecture which encourages active use of the objects and their metadata. In particular, an FDO has five parts: (i) The FDO *content*, bit sequences stored in an accessible repository; (ii) a *Persistent Identifier* (PID) such as a DOI that identifies the FDO and can resolve these same parts; (iii) Associated rich *metadata*, as separate FDOs; (iv) Type definitions, also separate FDOs; (v) Associated *operations* for the given types. A Digital Object typed as a Collection aggregates other DOs by reference.

The Digital Object Interface Protocol [107] can be considered an "abstract protocol" of requirements, DOs could be implemented in multiple ways. One suggested implementation is the FAIR Digital Object Framework[82], based on HTTP and the Linked Data Principles. While there is agreement on using PIDs based on DOIs, consensus on how to represent common metadata, core types and collections as FDOs has not yet been reached. We argue that RO-Crate can play an important role for FDOs:

1. By providing a predictable and extensible serialisation of structured metadata.
2. By formalising how to aggregate digital objects as collections (and adding their context).
3. By providing a natural Metadata FDO in the form of the RO-Crate Metadata File.
4. By being based on Linked Data and schema.org vocabulary, meaning that PIDs already exist for common types and properties.

At the same time, it is clear that the goal of FDO is broader than that of RO-Crate; namely, FDOs are active objects with distributed operations, and add further constraints such as PIDs for every element. These features improve FAIR features of digital objects and are also useful for RO-Crate, but they also severely restrict the infrastructure that needs to be implemented and maintained in order for FDOs to remain accessible. RO-Crate, on the other hand, is more flexible: it can minimally be used within any file system structure, or ideally exposed through a range of Web-based scenarios. A *FAIR profile of RO-Crate* (e.g. enforcing PID usage) will fit well within a FAIR Digital Object ecosystem.

---

[81]<https://w3id.org/ro/bagit>
[82]<https://fairdigitalobjectframework.org/>

## 5.3. Packaging Workflows

The use of computational workflows, typically combining a chain of tools in an analytical pipeline, has gained prominence in particular in the life sciences. Workflows might be used primarily to improve computational scalability, as well as to also assist in making computed data results FAIR [4], for instance by improving reproducibility [108], but also because programmatic data usage help propagate their metadata and provenance [109]. At the same time, workflows raise additional FAIR challenges, since they can be considered important research artefacts themselves. This viewpoint poses the problem of capturing and explaining the computational methods of a pipeline in sufficient machine-readable detail [3].

Even when researchers follow current best practices for workflow reproducibility [110] [108], the communication of computational outcomes through traditional academic publishing routes effectively adds barriers as authors are forced to rely on a textual manuscript representations. This hinder reproducibility and FAIR use of the knowledge previously captured in the workflow.

As a real-life example, let us look at a metagenomics article [111] that describes a computational pipeline. Here the authors have gone to extraordinary efforts to document the individual tools that have been reused, including their citations, versions, settings, parameters and combinations. The *Methods* section is two pages in tight double-columns with twenty four additional references, supported by the availability of data on an FTP server (60 GB) [112] and of open source code in GitHub Finn-Lab/MGS-gut[83] [113], including the pipeline as shell scripts and associated analysis scripts in R and Python.

This attention to reporting detail for computational workflows is unfortunately not yet the norm, and although bioinformatics journals have strong *data availability* requirements, they frequently do not require authors to include or cite *software, scripts and pipelines* used for analysing and producing results [114]. Indeed, in the absence of a specific requirement and an editorial policy to back it up – such as eliminating the reference limit – authors are effectively discouraged from properly and comprehensively citing software [115].

However detailed this additional information might be, another researcher who wants to reuse a particular computational method may first want to assess if the described tool or workflow is Re-runnable (executable at all), Repeatable (same results for original inputs on same platform), Reproducible (same results for original inputs with different platform or newer tools) and ultimately Reusable (similar results for different input data), Repurposable (reusing parts of the method for making a new method) or Replicable (rewriting the workflow following the method description) [116][117].

Following the textual description alone, researchers would be forced to jump straight to evaluate "Replicable" by rewriting the pipeline from scratch. This can be expensive and error-prone. They would firstly need to install all the software dependencies and download reference datasets. This can be a daunting task, which may have to be repeated multiple times as workflows typically are developed at small scale on desktop computers, scaled up to local clusters, and potentially put into production using cloud instances, each of which will have different requirements for software installations.

In recent years the situation has been greatly improved by software packaging and container technologies like Docker and Conda, these technologies have been increasingly adopted in life sciences

---

[83]<https://github.com/Finn-Lab/MGS-gut>

[118] thanks to collaborative efforts such as BioConda [119] and BioContainers [120], and support by Linux distributions (e.g. Debian Med [121]). As of November 2021, more than 9,000 software packages are available in BioConda alone[84], and 10,000 containers in BioContainers[85].

Docker and Conda have been integrated into workflow systems such as Snakemake [67], Galaxy [122] and Nextflow [66], meaning a downloaded workflow definition can now be executed on a "blank" machine (except for the workflow engine) with the underlying analytical tools installed on demand. Even with using containers there is a reproducibility challenge, for instance Docker Hub's retention policy will expire container images after six months[86], or a lack of recording versions of transitive dependencies of Conda packages could cause incompatibilities if the packages are subsequently updated.

These container and package systems only capture small amounts of metadata[87]. In particular, they do not capture any of the semantic relationships between their content. Understanding these relationships is made harder by the opaque wrapping of arbitrary tools with unclear functionality, licenses and attributions.

From this we see that computational workflows are themselves complex digital objects that need to be recorded not just as files, but in the context of their execution environment, dependencies and analytical purpose in research – as well as other metadata (e.g. version, license, attribution and identifiers).

It is important to note that having all these computational details in order to represent them in an RO-Crate is an ideal scenario – in practice there will always be gaps of knowledge, and exposing all provenance details automatically would require improvements to the data sources, workflow, workflow engine and its dependencies. RO-Crate can be seen as a flexible annotation mechanism for augmenting automatic workflow provenance. Additional metadata can be added manually, e.g. for sensitive clinical data that cannot be publicly exposed[88], or to cite software that lack persistent identifiers. This inline *FAIRifying* allows researchers to achieve "just enough FAIR" to explain their computational experiments.

## 6. Conclusion

RO-Crate has been established as an approach to packaging digital research artefacts with structured metadata. This approach assists developers and researchers to produce and consume FAIR archives of their research.

RO-Crate is formed by a set of best practice recommendations, developed by an open and broad community. These guidelines show how to use "just enough" Linked Data standards in a consistent way. The use of structured metadata with a rich base vocabulary can cover general-purpose contextual relations, with a Linked Data foundation that ensures extensibility to domain- and application-specific uses. We can therefore consider an RO-Crate not just as a structured data archive, but as a multimodal scholarly knowledge graph that can help "FAIRify" and combine metadata of existing resources.

---

[84]<https://anaconda.org/bioconda/>

[85]<https://biocontainers.pro/#/registry>

[86]<https://www.docker.com/blog/docker-hub-image-retention-policy-delayed-and-subscription-updates/>

[87]Docker and Conda can use *build recipes*, a set of commands that construct the container image through downloading and installing its requirements. However these recipes are effectively another piece of software code, which may itself decay and become difficult to rerun.

[88]FAIR principle A2: *Metadata are accessible, even when the data are no longer available.* [5]

The adoption of simple Web technologies in the RO-Crate specification has helped a rapid development of a wide variety of supporting open source tools and libraries. RO-Crate fits into the larger landscape of open scholarly communication and FAIR Digital Object infrastructure, and can be integrated into data repository platforms. RO-Crate can be applied as a data/metadata exchange mechanism, assist in long-term archival preservation of metadata and data, or simply used at a small scale by individual researchers. Thanks to its strong community support, new and improved profiles and tools are being continuously added to the RO-Crate tooling landscape, making it easier for adopters to find examples and support for their own use case.

*6.1. Strictness vs flexibility*

There is always a tradeoff between flexibility and strictness [123] when deciding on semantics of metadata models. Strict requirements make it easier for users and code to consume and populate a model, by reducing choices and having mandated "slots" to fill in. But such rigidity can also restrict richness and applicability of the model, as it in turn enforce the initial assumptions about what can be described.

RO-Crate attempts to strike a balance between these tensions, and provides a common metadata framework that encourages extensions. However, just like the RO-Crate specification can be thought of as a *core profile* of schema.org in JSON-LD, we cannot stress the importance of also establishing domain-specific RO-Crate profiles and conventions, as explored in sections 2.2.6 and 4. Specialization comes hand-in-hand with the principle of *graceful degradation*; RO-Crate applications and users are free to choose the semantic detail level they participate at, as long as they follow the common syntactic requirements.

## 7. Future Work

The direction of future RO-Crate work is determined by the community around it as a collaborative effort. We currently plan on further outreach, building training material (including a comprehensive entry-level tutorial) and maturing the reference implementation libraries. We will also collect and build examples of RO-Crate *consumption*, e.g. Jupyter Notebooks that query multiple crates using knowledge graphs. In addition, we are exploring ways to support some entity types requested by users, e.g. detailed workflow runs or container provenance, which do not have a good match in Schema.org. Such support could be added, for instance, by integrating other vocabularies or by having separated (but linked) metadata files.

Furthermore, we want to better understand how the community uses RO-Crate in practice and how it contrasts with other related efforts; this will help us to improve our specification and tools. By discovering commonalities in emerging usage (e.g. additional schema.org types), the community helps to reduce divergence that could otherwise occur with proliferation of further RO-Crate profiles. We plan to gather feedback via user studies, with the Linked Open Data community or as part of EOSC Bring-your-own-Data training events.

We operate in an open community where future and potential users of RO-Crate are actively welcomed to participate and contribute feedback and requirements. In addition we are targeting a wider audience through extensive outreach activities[89] and by initiating new connections. Recent contacts include American Geophysical Union (AGU) on Data Citation Reliquary [124], National Institute of Standards and Technology (NIST) on material science and InvenioRDM[90] used by the Zenodo data repository. New Horizon Europe projects adapting RO-Crate include BY-COVID[91], which aims to improve FAIR access to data on COVID-19 and other infectious diseases.

The main addition in the upcoming 1.2 release of the RO-Crate specifications will be the formalization of profiles[92] for different categories of crates. Additional entity types have been requested by users, e.g. workflow runs, business workflows, containers and software packages, tabular data structures; these are not always matched well with existing schema.org types but may benefit from other vocabularies or even separate metadata files, e.g. from Frictionless Data[93]. We will be further aligning with and collaborating with related research artefact description efforts like CodeMeta[94] for software metadata, Science-on-schema.org[95] [125] for datasets, FAIR Digital Objects[96] [74] and activities in EOSC task forces[97] including the EOSC Interoperability Framework [16].

## 8. Acknowledgements

This work has received funding from the European Commission's Horizon 2020 research and innovation programme for projects BioExcel-2[98] (H2020-INFRAEDI-2018-1 823830), IBISBA 1.0[99] (H2020-INFRAIA-2017-1-two-stage 730976), PREP-IBISBA[100] (H2020-INFRADEV-2019-2 871118), EOSC-Life[101] (H2020-INFRAEOSC-2018-2 824087), SyntheSys+[102] (H2020-INFRAIA-2018-1 823827). From the Horizon Europe Framework Programme this work has received funding for BY-COVID[103] (HORIZON-INFRA-2021-EMERGENCY-01 101046203).

Björn Grüning is supported by DataPLANT (NFDI 7/1 – 42077441[104]), part of the German National Research Data Infrastructure (NFDI), funded by the Deutsche Forschungsgemeinschaft (DFG).

Ana Trisovic is funded by the Alfred P. Sloan Foundation (grant number P-2020-13988)[105]. Harvard Data Commons is supported by an award from Harvard University Information Technology (HUIT).

---

[89]<https://www.researchobject.org/ro-crate/outreach.html>
[90]<https://inveniosoftware.org/products/rdm/>
[91]<https://by-covid.org/>
[92]<https://www.researchobject.org/ro-crate/1.2-DRAFT/profiles>
[93]<https://frictionlessdata.io/>
[94]<https://codemeta.github.io/>
[95]<https://science-on-schema.org/>
[96]<https://fairdo.org/>
[97]<https://www.eosc.eu/task-force-faq>
[98]<https://cordis.europa.eu/project/id/823830>
[99]<https://cordis.europa.eu/project/id/730976>
[100]<https://cordis.europa.eu/project/id/871118>
[101]<https://cordis.europa.eu/project/id/824087>
[102]<https://cordis.europa.eu/project/id/823827>
[103]<https://cordis.europa.eu/project/id/101046203>
[104]<https://gepris.dfg.de/gepris/projekt/442077441>
[105]<https://sloan.org/grant-detail/9555>

*8.1. Contributions*

Author contributions to this article and the RO-Crate projet according to the Contributor Roles Taxonomy CASRAI CrEDiT[106] [126]:

**Stian Soiland-Reyes** Conceptualization, Data curation, Formal Analysis, Funding acquisition, Investigation, Methodology, Project administration, Software, Visualization, Writing – original draft, Writing – review & editing
**Peter Sefton** Conceptualization, Investigation, Methodology, Project administration, Resources, Software, Writing – review & editing
**Mercè Crosas** Writing – review & editing
**Leyla Jael Castro** Methodology, Writing – review & editing
**Frederik Coppens** Writing – review & editing
**José M. Fernández** Methodology, Software, Writing – review & editing
**Daniel Garijo** Methodology, Writing – review & editing
**Björn Grüning** Writing – review & editing
**Marco La Rosa** Software, Methodology, Writing – review & editing
**Simone Leo** Software, Methodology, Writing – review & editing
**Eoghan Ó Carragáin** Investigation, Methodology, Project administration, Writing – review & editing
**Marc Portier** Methodology, Writing – review & editing
**Ana Trisovic** Software, Writing – review & editing
**RO-Crate Community** Investigation, Software, Validation, Writing – review & editing
**Paul Groth** Methodology, Supervision, Writing – original draft, Writing – review & editing
**Carole Goble** Conceptualization, Funding acquisition, Methodology, Project administration, Supervision, Visualization, Writing – review & editing

We would also like to acknowledge contributions from:

**Finn Bacall** Software, Methodology
**Herbert Van de Sompel** Writing – review & editing
**Ignacio Eguinoa** Software, Methodology
**Nick Juty** Writing – review & editing
**Oscar Corcho** Writing – review & editing
**Stuart Owen** Writing – review & editing
**Laura Rodríguez-Navas** Software, Visualization, Writing – review & editing
**Alan R. Williams** Writing – review & editing

---

[106]<https://casrai.org/credit/>

# Appendix A. Formalizing RO-Crate in First Order Logic

Below is a formalization of the concept of RO-Crate as a set of relations using First Order Logic:

## A.1. Language

Definition of language $\mathcal{L}_{rocrate}$:

$$\mathcal{L}_{rocrate} = \{Property(p), Class(c), Value(x), \mathbb{R}, \mathbb{S}\}$$
$$\mathbb{D} = \mathbb{IRI}$$
$$\mathbb{IRI} \equiv \text{IRIs as defined in RFC3987}$$
$$\mathbb{R} \equiv \text{real or integer numbers}$$
$$\mathbb{S} \equiv \text{literal strings}$$

The domain of discourse $\mathbb{D}$ is the set of $\mathbb{IRI}$ identifiers [30] (notation `<http://example.com/>`)[107], with additional descriptions using numbers $\mathbb{R}$ (notation 13.37) and literal strings $\mathbb{S}$ (notation "Hello").

From this formalised language $\mathcal{L}_{rocrate}$ we can interpret an RO-Crate in any representation that can gather these descriptions, their properties, classes, and literal attributes.

## A.2. Minimal RO-Crate

Below we use $\mathcal{L}_{rocrate}$ to define a minimal[108] RO-Crate:

$$\begin{aligned}
ROCrate(R) \vDash\ & Root(R) \land Mentions(R,R) \land hasPart(R,d) \land \\
& Mentions(R,d) \land DataEntity(d) \land \\
& Mentions(R,c) \land ContextualEntity(c) \\
\forall r\ Root(r) \Rightarrow\ & Dataset(r) \land name(r,n) \land description(r,d) \land \\
& datePublished(r, date) \land license(e, l) \\
\forall e \forall n\ name(e,n) \Rightarrow\ & Value(n) \\
\forall e \forall s\ description(e,s) \Rightarrow\ & Value(s) \\
\forall e \forall d\ datePublished(e,d) \Rightarrow\ & Value(d) \\
\forall e \forall l\ license(e,l) \Rightarrow\ & ContextualEntity(l) \\
DataEntity(e) \equiv\ & File(e) \oplus Dataset(e)
\end{aligned}$$

---

[107] For simplicity, blank nodes are not included in this formalisation, as RO-Crate recommends the use of IRI identifiers: https://www.researchobject.org/ro-crate/1.1/appendix/jsonld.html#describing-entities-in-json-ld

[108] The full list of types, relations and attribute properties from the RO-Crate specification are not included. Examples shown include $datePublished$, $CreativeWork$ and $name$.

$$Entity(e) \equiv DataEntity(e) \lor ContextualEntity(e)$$

$$\forall e\, Entity(e) \Rightarrow type(e,c) \land Class(c)$$

$$\forall e\, ContextualEntity(e) \Rightarrow name(e,n)$$

$$Mentions(R,s) \vDash Relation(s,p,e) \oplus Attribute(s,p,l)$$

$$Relation(s,p,o) \vDash Entity(s) \land Property(p) \land Entity(o)$$

$$Attribute(s,p,v) \vDash Entity(s) \land Property(p) \land Value(v)$$

$$Value(v) \equiv v \in \mathbb{R} \oplus v \in \mathbb{S}$$

An $ROCrate(R)$ is defined as a self-described *Root Data Entity*, which describes and contains parts (*data entities*), which are further described in *contextual entities*. These terms align with their use in the RO-Crate 1.1 terminology[109].

The $Root(r)$ is a type of $Dataset(r)$, and must have the metadata to literal attributes to provide a *name*, *description* and *datePublished*, as well as a contextual entity identifying its license. These predicates correspond to the RO-Crate 1.1 requirements for the root data entity[110].

The concept of an $Entity(e)$ is introduced as being either a $DataEntity(e)$, a $ContextualEntity(e)$, or both[111]. Any $Entity(e)$ must be typed with at least one $Class(c)$, and every $ContextualEntity(e)$ must also have a $name(e,n)$; this corresponds to expectations for any *referenced contextual entity* (section 2.2.3).

For simplicity in this formalization (and to assist production rules below) $R$ is a constant representing a single RO-Crate, typically written to independent RO-Crate Metadata files. $R$ is used by $Mentions(R,e)$ to indicate that $e$ is an Entity described by the RO-Crate and therefore its metadata (a set of *Relation* and *Attribute* predicates) form part of the RO-Crate serialization. $Relation(s,p,o)$ and $Attribute(s,p,x)$ are defined as a *subject—predicate—object* triple pattern from an $Entity(s)$ using a $Property(p)$ to either another $Entity(o)$ or a $Value(x)$ value.

*A.3. Example of formalized RO-Crate*

The below is an example RO-Crate represented using the above formalisation, assuming a base URI of `<http://example.com/ro/123/>`:

$ROCrate$(`<http://example.com/ro/123/>`)

$name$(`<http://example.com/ro/123/>`,

"Data files associated with the manuscript:Effects of ...")

$description$(`<http://example.com/ro/123/`,

"Palliative care planning for nursing home residents ...")

$license$(`<http://example.com/ro/123/>`,

---

[109] https://www.researchobject.org/ro-crate/1.1/terminology
[110] https://www.researchobject.org/ro-crate/1.1/root-data-entity.html#direct-properties-of-the-root-data-entity
[111] https://www.researchobject.org/ro-crate/1.1/contextual-entities.html#contextual-vs-data-entities

    <https://spdx.org/licenses/CC-BY-4.0>)

*datePublished*(<http://example.com/ro/123/>, "2017-02-23")

*hasPart*(<http://example.com/ro/123/>, <http://example.com/ro/123/file.txt>)

*hasPart*(<http://example.com/ro/123/>, <http://example.com/ro/123/interviews/>)

*ContextualEntity*(<https://spdx.org/licenses/CC-BY-4.0>)

*name*(<https://spdx.org/licenses/CC-BY-4.0>,
   Creative Commons Attribution 4.0")

*ContextualEntity*(<https://spdx.org/licenses/CC-BY-NC-4.0>)

*name*(<https://spdx.org/licenses/CC-BY-NC-4.0>,
   Creative Commons Attribution Non Commercial 4.0")

*File*(<http://example.com/ro/123/survey.csv>)

*name*(<http://example.com/ro/123/survey.csv>, "Survey of care providers")

*Dataset*(<http://example.com/ro/123/interviews/>)

*name*(<http://example.com/ro/123/interviews/>,
   "Audio recordings of care provider interviews")

*license*(<http://example.com/ro/123/interviews/>,
   <https://spdx.org/licenses/CC-BY-NC-4.0>)

    Notable from this triple-like formalization is that a RO-Crate *R* is fully represented as a tree at depth 2 helped by the use of Iri nodes. For instance the aggregation from the root entity *hasPart*(…`interviews/>`) is at same level as the data entity's property *license*(…`CC-BY-NC-4.0>`) and that contextual entity's attribute *name*(…Non Commercial 4.0"). As shown in section 2.3.1, the RO-Crate Metadata File serialization is an equivalent shallow tree, although at depth 3 to cater for the JSON-LD preamble of `"@context"` and `"@graph"`.

    In reality many additional attributes and contextual types from schema.org types like http://schema.org/affiliation and http://schema.org/Organization would be used to further describe the RO-Crate and its entities, but as these are optional (*SHOULD* requirements) they do not form part of this formalization.

## A.4. Mapping to RDF with schema.org

A formalized RO-Crate in $\mathcal{L}_{rocrate}$ can be mapped to different serializations. Assume a simplified[112] language $\mathcal{L}_{RDF}$ based on the RDF abstract syntax [127]:

$$\mathcal{L}_{RDF} \equiv \{Triple(s, p, o), IRI(i), BlankNode(b), Literal(s), \mathbb{IRI}, \mathbb{S}, \mathbb{R}\}$$

$$\mathbb{D}_{RDF} \equiv \mathbb{S}$$

$$\forall i\, IRI(i) \Rightarrow i \in \mathbb{IRI}$$

$$\forall s \forall p \forall o\, Triple(s, p, o) \Rightarrow \bigl(IRI(s) \vee BlankNode(s)\bigr) \wedge$$

$$IRI(p) \wedge$$

$$\bigl(IRI(o) \vee BlankNode(o) \vee Literal(o)\bigr)$$

$$Literal(v) \vDash Value(v) \wedge Datatype(v, t) \wedge IRI(t)$$

$$\forall v\, Value(v) \Rightarrow v \in \mathbb{S}$$

$$LanguageTag(v, l) \equiv Datatype\bigl(v,$$

$$\texttt{<http://www.w3.org/1999/02/22-rdf-syntax-ns\#langString>}\bigr)$$

Below follows a mapping from $\mathcal{L}_{rocrate}$ to $\mathcal{L}_{RDF}$ using schema.org as vocabulary:

$$Property(p) \Rightarrow type(p, \texttt{<http://www.w3.org/2000/01/rdf-schema\#Property>})$$

$$Class(c) \Rightarrow type(c, \texttt{<http://www.w3.org/2000/01/rdf-schema\#Class>})$$

$$Dataset(d) \Rightarrow type(d, \texttt{<http://schema.org/Dataset>})$$

$$File(f) \Rightarrow type(f, \texttt{<http://schema.org/MediaObject>})$$

$$ContextualEntity(e) \Rightarrow type(f, \texttt{<http://schema.org/Thing>})$$

$$CreativeWork(e) \Rightarrow ContextualEntity(e) \wedge type(e, \texttt{<http://schema.org/CreativeWork>})$$

$$hasPart(e, t) \Rightarrow Relation(e, \texttt{<http://schema.org/hasPart>}, t)$$

$$name(e, n) \Rightarrow Attribute(e, \texttt{<http://schema.org/name>}, n)$$

$$description(e, s) \Rightarrow Attribute(e, \texttt{<http://schema.org/description>}, s)$$

$$datePublished(e, d) \Rightarrow Attribute(e, \texttt{<http://schema.org/datePublished>}, d)$$

$$license(e, l) \Rightarrow Relation(e, \texttt{<http://schema.org/license>}, l) \wedge CreativeWork(l)$$

$$type(e, t) \Rightarrow Relation(e, \texttt{<http://www.w3.org/1999/02/22-rdf-syntax-ns\#type>}, t)$$

$$\wedge Class(t)$$

---

[112]This simplification does not cover the extensive list of literal datatypes built-in to RDF 1.1, only strings and decimal real numbers. Likewise, language of literals are not included.

$$String(s) \equiv Value(s) \land s \in \mathbb{S}$$
$$String(s) \Rightarrow Datatype(s, \text{<http://www.w3.org/2001/XMLSchema\#string>})$$
$$Decimal(d) \equiv Value(d) \land d \in \mathbb{R}$$
$$Decimal(d) \Rightarrow Datatype(d, \text{<http://www.w3.org/2001/XMLSchema\#decimal>})$$
$$Relation(s, p, o) \Rightarrow Triple(s, p, o) \land IRI(s) \land IRI(o)$$
$$Attribute(s, p, o) \Rightarrow Triple(s, p, o) \land IRI(s) \land Literal(o)$$

Note that in the JSON-LD serialization of RO-Crate, the expression of *Class* and *Property* is typically indirect: The JSON-LD `@context` maps to schema.org IRIs, which, when resolved as Linked Data, embed their formal definition as RDFa. Extensions may however include such term definitions directly in the RO-Crate.

*A.5. RO-Crate 1.1 Metadata File Descriptor*

An important RO-Crate principle is that of being **self-described** Therefore the serialisation of the RO-Crate into a file should also describe itself in a Metadata File Descriptor[113], indicating it is *about* (describing) the RO-Crate root data entity, and that it *conformsTo* a particular version of the RO-Crate specification:

$$about(s, o) \Rightarrow Relation(s, \text{<http://schema.org/about>}, o)$$
$$conformsTo(s, o) \Rightarrow Relation(s, \text{<http://purl.org/dc/terms/conformsTo>}, o)$$
$$MetadataFile(m) \Rightarrow CreativeWork(m) \land about(m, R) \land ROCrate(R) \land$$
$$conformsTo(m, \text{<https://w3id.org/ro/crate/1.1>})$$

Note that although the metadata file necessarily is an *information resource* written to disk or served over the network (as JSON-LD), it is not considered to be a contained *part* of the RO-Crate in the form of a *data entity*, rather it is described only as a *contextual entity*.

In the conceptual model the *RO-Crate Metadata File* can be seen as the top-level node that describes the *RO-Crate Root*, however in the formal model (and the JSON-LD format) the metadata file descriptor is an additional contextual entity that is not affecting the depth-limit of the RO-Crate.

*A.6. Forward-chained Production Rules for JSON-LD*

Combining the above predicates and schema.org mapping with rudimentary JSON templates, these forward-chaining production rules can output JSON-LD according to the RO-Crate 1.1 specification[114]:

---

[113]https://www.researchobject.org/ro-crate/1.1/root-data-entity.html#ro-crate-metadata-file-descriptor

[114]**Limitations:** Contextual entities not related from the RO-Crate (e.g. using inverse relations to a data entity) would not be covered by the single direction $Mentions(R, s)$ production rule; see GitHub issue ResearchObject/ro-crate#122. The $datePublished(e, d)$ rule do not include syntax checks for the ISO 8601 datetime format. Compared with RO-Crate

$$Mentions(R, s) \land Relation(s, p, o) \Rightarrow Mentions(R, o)$$

$$IRI(i) \Rightarrow \texttt{"}i\texttt{"}$$

$$Decimal(d) \Rightarrow d$$

$$String(s) \Rightarrow \texttt{"}s\texttt{"}$$

$$\forall e \forall t\; type(e, t) \Rightarrow \{\texttt{"@id"}{:}e,$$
$$\texttt{"@type"}{:}t$$
$$\}$$

$$\forall s \forall p \forall o\; Relation(s, p, o) \Rightarrow \{\texttt{"@id"}{:}s,$$
$$p{:}\ \{\ \texttt{"@id"}{:}o\}$$
$$\}$$

$$\forall s \forall p \forall v\; Attribute(s, p, v) \Rightarrow \{\texttt{"@id"}{:}s,$$
$$p{:}v$$
$$\}$$

$$\forall r \forall c ROCrate(r) \Rightarrow \{\ \texttt{"@graph"}{:}\ [$$
$$Mentions(r, c)\ *$$
$$]$$
$$\}$$

$$R \equiv \texttt{<./>}$$

$$R \Rightarrow MetadataFile(\texttt{<ro-crate-metadata.json>})$$

This exposes the first order logic domain of discourse of IRIs, with rational numbers and strings as their corresponding JSON-LD representation. These production rules first grow the graph of $R$ by adding a transitive rule – anything described in $R$ which is related to $o$, means that $o$ is also mentioned by the $ROCrate(R)$. For simplicity this rule is one-way; in theory the graph can also contain free-standing contextual entities that have outgoing relations to data- and contextual entities, but these are proposed to be bound to the root data entity with schema.org relation .

---

examples, this generated JSON-LD does not use a $@context$ as the IRIs are produced unshortened, a post-step could do JSON-LD Flattening with a versioned RO-Crate context. The `@type` expansion is included for clarity, even though this is also implied by the $type(e, t)$ expansion to $Relation(e, \texttt{xsd:type})$.

**Appendix B.  RO-Crate Community**

As of 2021-10-04, the *RO-Crate* Community members are:
- Peter Sefton https://orcid.org/0000-0002-3545-944X (co-chair)
- Stian Soiland-Reyes https://orcid.org/0000-0001-9842-9718 (co-chair)
- Eoghan Ó Carragáin https://orcid.org/0000-0001-8131-2150 (emeritus chair)
- Oscar Corcho https://orcid.org/0000-0002-9260-0753
- Daniel Garijo https://orcid.org/0000-0003-0454-7145
- Raul Palma https://orcid.org/0000-0003-4289-4922
- Frederik Coppens https://orcid.org/0000-0001-6565-5145
- Carole Goble https://orcid.org/0000-0003-1219-2137
- José María Fernández https://orcid.org/0000-0002-4806-5140
- Kyle Chard https://orcid.org/0000-0002-7370-4805
- Jose Manuel Gomez-Perez https://orcid.org/0000-0002-5491-6431
- Michael R Crusoe https://orcid.org/0000-0002-2961-9670
- Ignacio Eguinoa https://orcid.org/0000-0002-6190-122X
- Nick Juty https://orcid.org/0000-0002-2036-8350
- Kristi Holmes https://orcid.org/0000-0001-8420-5254
- Jason A. Clark https://orcid.org/0000-0002-3588-6257
- Salvador Capella-Gutierrez https://orcid.org/0000-0002-0309-604X
- Alasdair J. G. Gray https://orcid.org/0000-0002-5711-4872
- Stuart Owen https://orcid.org/0000-0003-2130-0865
- Alan R Williams https://orcid.org/0000-0003-3156-2105
- Giacomo Tartari https://orcid.org/0000-0003-1130-2154
- Finn Bacall https://orcid.org/0000-0002-0048-3300
- Thomas Thelen https://orcid.org/0000-0002-1756-2128
- Hervé Ménager https://orcid.org/0000-0002-7552-1009
- Laura Rodríguez-Navas https://orcid.org/0000-0003-4929-1219
- Paul Walk https://orcid.org/0000-0003-1541-5631
- brandon whitehead https://orcid.org/0000-0002-0337-8610
- Mark Wilkinson https://orcid.org/0000-0001-6960-357X
- Paul Groth https://orcid.org/0000-0003-0183-6910
- Erich Bremer https://orcid.org/0000-0003-0223-1059
- LJ Garcia Castro https://orcid.org/0000-0003-3986-0510
- Karl Sebby https://orcid.org/0000-0001-6022-9825
- Alexander Kanitz https://orcid.org/0000-0002-3468-0652
- Ana Trisovic https://orcid.org/0000-0003-1991-0533
- Gavin Kennedy https://orcid.org/0000-0003-3910-0474
- Mark Graves https://orcid.org/0000-0003-3486-8193
- Jasper Koehorst https://orcid.org/0000-0001-8172-8981
- Simone Leo https://orcid.org/0000-0001-8271-5429
- Marc Portier https://orcid.org/0000-0002-9648-6484
- Paul Brack https://orcid.org/0000-0002-5432-2748
- Milan Ojsteršek https://orcid.org/0000-0003-1743-8300
- Bert Droesbeke https://orcid.org/0000-0003-0522-5674

- Chenxu Niu https://github.com/UstcChenxu
- Kosuke Tanabe https://orcid.org/0000-0002-9986-7223
- Tomasz Miksa https://orcid.org/0000-0002-4929-7875
- Marco La Rosa https://orcid.org/0000-0001-5383-6993
- Cedric Decruw https://orcid.org/0000-0001-6387-5988
- Andreas Czerniak https://orcid.org/0000-0003-3883-4169
- Jeremy Jay https://orcid.org/0000-0002-5761-7533
- Sergio Serra https://orcid.org/0000-0002-0792-8157
- Ronald Siebes https://orcid.org/0000-0001-8772-7904
- Shaun de Witt https://orcid.org/0000-0003-4196-3658
- Shady El Damaty https://orcid.org/0000-0002-2318-4477
- Douglas Lowe https://orcid.org/0000-0002-1248-3594
- Sergio Serra https://orcid.org/0000-0002-0792-8157
- Xuanqi Li https://orcid.org/0000-0003-1498-6205